\documentclass[brazilian,twocolumn,prb,superscriptaddress,amsmath,amssymb,floatfix,preprintnumbers]{revtex4-2}
\usepackage[latin9,utf8]{inputenc}
\usepackage{float}
\usepackage{textcomp}
\usepackage{amsmath}
\usepackage{graphicx}
\makeatletter
\usepackage{url}
\usepackage{grffile}
\usepackage{bbold}

\newcommand{\lyxmathsym}[1]{\ifmmode\begingroup\def\b@ld{bold}
  \text{\ifx\math@version\b@ld\bfseries\fi#1}\endgroup\else#1\fi}

\@ifundefined{textcolor}{}
{%
 \definecolor{BLACK}{gray}{0}
 \definecolor{WHITE}{gray}{1}
 \definecolor{RED}{rgb}{1,0,0}
 \definecolor{GREEN}{rgb}{0,1,0}
 \definecolor{BLUE}{rgb}{0,0,1}
 \definecolor{CYAN}{cmyk}{1,0,0,0}
 \definecolor{MAGENTA}{cmyk}{0,1,0,0}
 \definecolor{YELLOW}{cmyk}{0,0,1,0}
}

\usepackage{dcolumn}
\usepackage{color}
\DeclareGraphicsExtensions{.png .jpg .pdf}

\usepackage{hyperref}
\hypersetup{
     colorlinks = true,
     linkcolor = blue,
     anchorcolor = blue,
     citecolor = blue,
     filecolor = blue,
     urlcolor = blue
     }
\usepackage{braket}
\usepackage{physics}
\usepackage{array}
\usepackage{booktabs}
\usepackage{soul}

\makeatother

\begin{document}

\title{Effect of \textit{zitterbewegung} on the propagation of wave packets in ABC-stacked multilayer graphene: an analytical and computational approach}

\author{I. R. Lavor}
\email{icaro@fisica.ufc.br}

\affiliation{Instituto Federal de Educação, Ciência e Tecnologia do Maranhão,
KM-04, Enseada, 65200-000, Pinheiro, Maranhão, Brazil}

\affiliation{Departamento de F\'{i}sica, Universidade Federal do Ceará, Caixa
Postal 6030, Campus do Pici, 60455-900 Fortaleza, Ceará, Brazil}

\affiliation{Department of Physics, University of Antwerp, Groenenborgerlaan 171, B-2020 Antwerp, Belgium}

\author{D. R. da Costa}
\email{diego\_rabelo@fisica.ufc.br}
\affiliation{Departamento de F\'{i}sica, Universidade Federal do Ceará, Caixa
Postal 6030, Campus do Pici, 60455-900 Fortaleza, Ceará, Brazil}

\author{Andrey Chaves}
\email{andrey@fisica.ufc.br}

\affiliation{Departamento de F\'{i}sica, Universidade Federal do Ceará, Caixa
Postal 6030, Campus do Pici, 60455-900 Fortaleza, Ceará, Brazil}
\affiliation{Department of Physics, University of Antwerp, Groenenborgerlaan 171,
B-2020 Antwerp, Belgium}

\author{S. H. R. Sena}

\affiliation{Instituto de Ciências Exatas e da Natureza, Universidade da Integração Internacional da Lusofonia Afro-Brasileira, Centro, 62790-000 Redenção, Ceará, Brasil}

\author{G. A. Farias}

\affiliation{Departamento de F\'{i}sica, Universidade Federal do Ceará, Caixa
Postal 6030, Campus do Pici, 60455-900 Fortaleza, Ceará, Brazil}

\author{B. Van Duppen}

\affiliation{Department of Physics, University of Antwerp, Groenenborgerlaan 171,
B-2020 Antwerp, Belgium}

\author{F. M. Peeters}
\email{francois.peeters@uantwerpen.be}

\affiliation{Department of Physics, University of Antwerp, Groenenborgerlaan 171,
B-2020 Antwerp, Belgium}

\date{\today }
\begin{abstract}
The time evolution of a low-energy two-dimensional Gaussian wave packet in ABC-stacked $n$-layer graphene (ABC-NLG) is investigated. Expectation values of the position $(x,y)$ of center-of-mass and the total probability densities of the wave packet are calculated analytically using the Green's function method. These results are confirmed using an alternative numerical method based on the split-operator technique within the Dirac approach for ABC-NLG, which additionally allows to include external fields and potentials. The main features of the \textit{zitterbewegung} (trembling motion) of wave packets in graphene are demonstrated and are found to depend not only on the wave packet width and initial pseudospin polarization, but also on the number of layers. Moreover, the analytical and numerical methods proposed here allow to investigate wave packet dynamics in graphene systems with an arbitrary number of layers and arbitrary potential landscapes.
\end{abstract}

\maketitle

\section{Introduction\label{sec:Introduction}}

\textit{Zitterbewegung} (ZBW) is a fast oscillation or trembling motion of elementary particles that obey the Dirac equation\citep{Dirac1928}, which was predicted by Erwin Schr{\"o}dinger in $1930$ for relativistic fermions\citep{Schroedinger1930a}. Schr{\"o}dinger observed that the component of relativistic velocity for electrons in vacuum does not commute with the free-electron Hamiltonian. Consequently, the expectation value of the position of these electrons displays rapid oscillatory motion, owing to the fact that the velocity is not a constant of motion. It was also demonstrated that ZBW occurs due to the interference between the positive and negative energy states in the wave packet, and the characteristic frequency of this motion is determined by the gap between the two states. 

In the last decades, Schr{\"o}dinger's idea stimulated numerous theoretical studies e.g. in ultracold atoms\citep{Vaishnav2008,Merkl2008}, semiconductors\citep{Schliemann2005,Zawadzki2005,Schliemann2006,Rusin2007a,Schliemann2008,Biswas2014}, carbon nanotubes\citep{Zawadzki2006}, topological insulators\citep{Shi2013}, crystalline solids\citep{Ferrari1990,Zawadzki2010} and other systems\citep{Cannata1991,Vonsovskii1993,Lamata2007,Cunha2019}. Although ZBW was theoretically found using a quantum simulation of the Dirac equation for trapped ions\citep{Gerritsma2010}, Bose--Einstein condensates\citep{Wang2010,LeBlanc2013,Qu2013} and, most recently, an optical simulation\citep{Silva2019}, up to now, no direct experimental observations have been carried out. The reason is that the Dirac equation predicts ZBW with amplitude of the order of the Compton wavelength ($10^{-2}\ \lyxmathsym{\AA}$) and a frequency of $\omega_{ZB}\approx10^{21}\ \text{Hz}$, which are not accessible with current experimental techniques.

With the discovery of graphene\citep{Novoselov2004,Novoselov2005}, a single-layer of a honeycomb lattice of carbon atoms with unique electronic properties\citep{Neto2009,McCann2013,Choi2019,Katsnelson2006,Novoselov2005,Wallace1947,McCann2006,Pereira2010}, the ZBW effect has been revisited recently\citep{Cserti2006,Rusin2007,Trauzettel2007,Maksimova2008,Rusin2008,A.K.2009,Wang2010,Deng2015,serna2019pseudospin}, since low-energy electrons in graphene behave as quasi-relativistic particles\citep{Kim2017,Avouris2010,Katsnelson2007}. Maksimova \textit{et al}.\citep{Maksimova2008} investigated the wave packet evolution in monolayer graphene (MLG) analytically for different pseudo-spin polarizations using the Green's function method. Rusin and Zawadzki\citep{Rusin2007} analyzed the evolution of a Gaussian wave packet in MLG and bilayer graphene (BLG), as well as in carbon nanotubes, for one kind of initial pseudo-spin polarization, which is directly linked to the direction of propagation of the wave packet. They demonstrated that the transient character of ZBW in BLG is related to the movement in opposite directions of the sub-wave packets corresponding to the positive and negative energy contributions. A similar investigation for MLG was performed pure numerically based on the so-called split-operator technique (SOT), which will be explained more in details later one here, by Chaves \textit{et al}.\citep{Chaves2010}, and, most recently, in multilayer phosphorene by Cunha \textit{et al}.\cite{Cunha2019}, that compared both SOT and Green's function results.

In this paper, we generalize the previous studies on ZBW in MLG by proposing different techniques to study the dynamics of charged particles described by a two-dimensional (2D) Gaussian wave packet in $ABC$ stacked $n-$layer graphene ($ABC$-NLG). We use an approximated $2\times2$ Hamiltonian valid for low-energy electrons in $ABC$-NLG and the Green's function formalism to obtain the time-evolved electron wave function for an arbitrary pseudospin polarization and then use this result to analytically calculate the expectation values of center-of-mass coordinates, the trajectory and spreading of the wave packet in real space, as well as their oscillations due to ZBW. We also develop a numerical method to perform the same calculation based on the SOT, but with much higher flexibility, allowing to consider $ABC$-NLG and any potential profile. Results from both theoretical approaches for MLG, BLG and trilayer graphene (TLG) are compared and their validity is verified. The dependence of several qualitative features of ZBW on the number of graphene layers and wave packet initial conditions is discussed in detail. The analytical and numerical methods proposed here can be straightforwardly adapted to investigate transport properties of multi-layer graphene in the presence of external fields and arbitrary potential profiles.

\begin{figure}
\centering{\includegraphics[width=1\columnwidth]{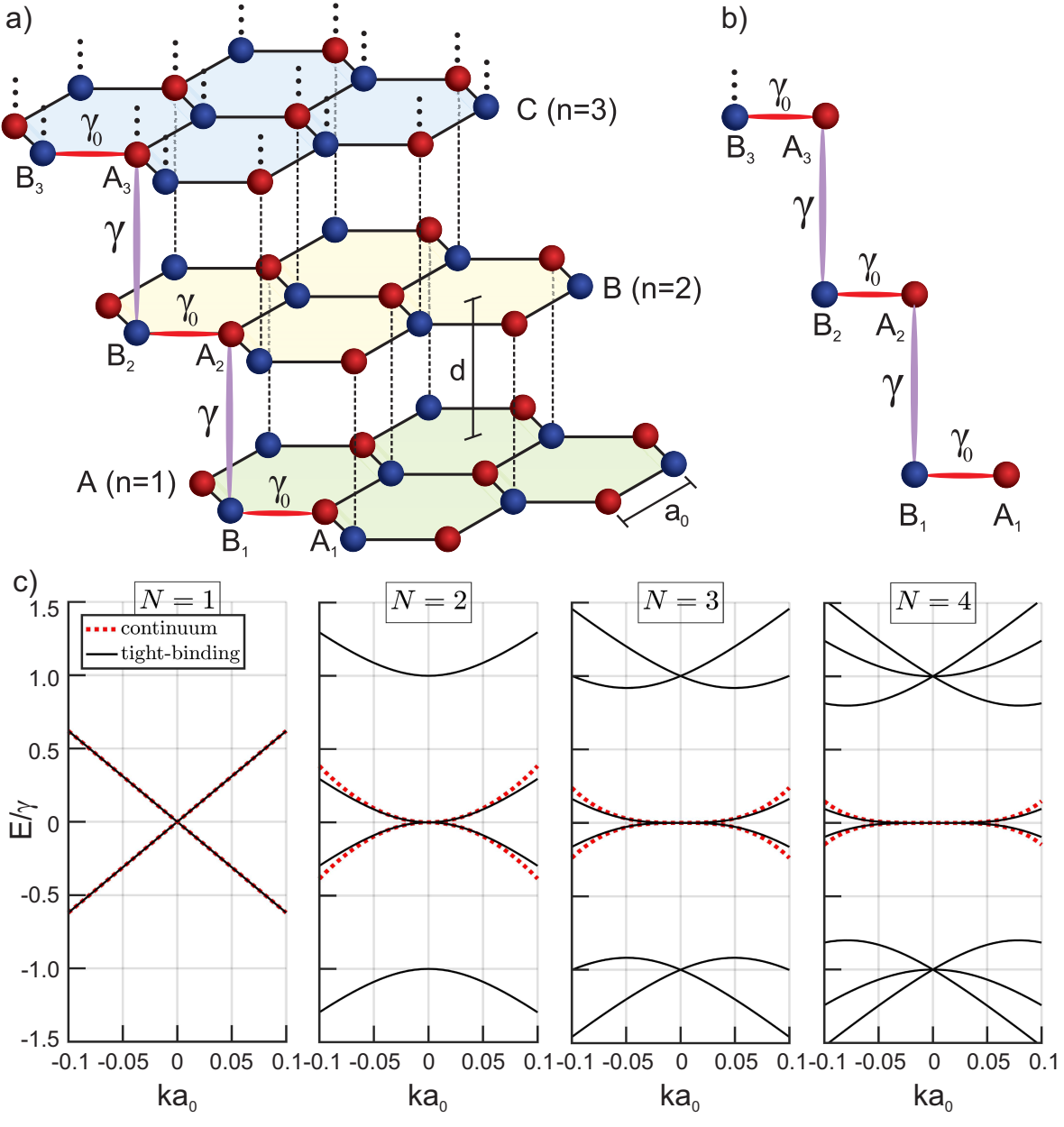}}
\caption{(Color online) (a) Schematic representation for NLG with rhombohedral stacking (ABC). The interlayer and intralayer distance are $d\approx3.35\ \mathring{\text{A}}$ and $a_0=1.42\ \mathring{\text{A}}$, respectively. The two non-equivalent carbon sublattices in each layer are indicated by red ($A$) and blue ($B$) circular symbols. (b) Representation of ABC-stacked multi-layer graphene with intralayer hopping between first nearest neighbors $\gamma_{0}$ and interlayer hopping energy between $A_{i}$ and $B_{i+1}$ sites of each layer given by $\gamma$. (c) Energy spectrum of multilayer graphene near one of the Dirac cones for low energies obtained by tight-binding model (solid black curves) and two-band continuum model (red dashed curves). The energy is expressed in units of the interlayer hopping energy $\gamma$ and the wave vector is expressed in units of $a_0^{-1}$, the inverse of the nearest-neighbour interatomic distance.}
\label{Fig: Geometries-1} 
\end{figure}

\section{The base of ZBW in $N$-ABC-stacked multilayer graphene\label{sec:THEORETICAL-MODEL}}

For $ABC$-NLG, as illustrated in Fig.~\ref{Fig: Geometries-1}(a), the effective Hamiltonian in the long wavelength approximation, near the $K$ point on the first Brillouin zone of $n$ graphene layers, can be written as the following approximated $2n\times2n$ matrix\citep{Duppen2013}
\begin{equation}
H_{n}=\hbar v_{F}\left[\begin{array}{ccccc}
\vec{\sigma}\cdot\vec{k} & \tau & 0 & \cdots & 0\\
\tau^{\dagger} & \vec{\sigma}\cdot\vec{k} & \tau & \cdots & 0\\
0 & \tau^{\dagger} & \vec{\sigma}\cdot\vec{k} & \ddots & 0\\
\vdots & \vdots & \ddots & \ddots & \tau\\
0 & 0 & 0 & \tau^{\dagger} & \vec{\sigma}\cdot\vec{k}
\end{array}\right] +\mathcal{V}\mathbb{1},\label{eq:hamiltonian-n layers}
\end{equation}
by considering only nearest-neighbor interlayer transitions, being $\tau$ represented the $2\times 2$ coupling matrix given by
\begin{equation}
\tau=\frac{1}{\hbar v_{F}}\left[\begin{array}{cc}
0 & 0\\
\gamma & 0
\end{array}\right],
\end{equation}
with $\gamma\approx 0.4\ \text{eV}$ being the interlayer hopping parameter\citep{Partoens2006}, as shown in Fig.~\ref{Fig: Geometries-1}(b). $v_F = 3a_0\gamma_0/2\hbar$ is the Fermi velocity with $\gamma_0\approx 2.7\ \text{eV}$ being the intralayer coupling, $\vec{\sigma}=\left(\sigma_{x},\sigma_{y},\sigma_{z}\right)$ are the Pauli matrices and $\vec{k}=\left(k_{x},k_{y}\right)$ is the wave vector. Note that the tridiagonal matrix, Eq.~(\ref{eq:hamiltonian-n layers}), only considers the coupling between the adjacent layers, otherwise off-tridiagonal terms would be non-zero, and its main diagonal is composed by $n$ MLG-type Hamiltonians. Within a low-energy approximation ($\left|E\right|\ll\gamma$), it is possible to rewrite Eq. (\ref{eq:hamiltonian-n layers}) as an effective two-band Hamiltonian\citep{Nakamura2008,Manes2007,Kumar2012}
\begin{equation}
H_{n}\left(k\right)=\frac{\left(\hbar v_{F}k\right)^{n}}{\gamma^{n-1}}\left[\begin{array}{cc}
0 & e^{-in\phi}\\
e^{in\phi} & 0
\end{array}\right]+\mathcal{V}\mathbb{1},
\label{eq:hamiltonian n-ABC}
\end{equation}
where $\phi=\arctan\left(k_{y}/k_{x}\right)$ is the 2D polar angle in momentum space, and the eigenstate that was given by a $2n-$component wave function $\Psi^{n}=\left(\Psi_{A}^{1},\Psi_{B}^{1},\Psi_{A}^{2},\Psi_{B}^{2}\cdots\Psi_{A}^{n}\Psi_{B}^{n}\right)$ is now approximated by the two-component one $\Psi^{n}\rightarrow\Psi_{\text{eff}}=\left[\Psi_{A}^{1}\quad \Psi_{B}^{n} \right]^T$.\citep{Prarokijjak2015,McCann2007} An arbitrary external electric potential, e.g. a perpendicular electric field, can be incorporated in the model by adding a potential energy $\mathcal{V}_i$ to the on-site energies in the main diagonal, with $i = 1$, $2$, $\cdots n$ and $n$ being the number of layers, as represent by the second term $\mathcal{V}\mathbb{1}$ in Eqs.~(\ref{eq:hamiltonian-n layers}) and (\ref{eq:hamiltonian n-ABC}), where $\mathbb{1}$ denotes the identity matrix with dimension $2n\times 2n$ and $2\times 2$, respectively. The only assumption to this approach of adding an external potential in the two-band model is that the field affects equally the on-site energies of all atoms in the same layer $i$, and only the potential difference between the first and last layers is taken into account. For the sake of simplicity but without loss of generality, we assumed in the present paper that the multilayer graphene system is free of interactions with any external sources. The low-energy bands described by this effective two-band Hamiltonian (\ref{eq:hamiltonian n-ABC}) arise from hopping between the non-dimer sites, as can be illustrated for instance in Fig.~\ref{Fig: Geometries-1} by the coupling between $A_1$ and $B_2$ sites and $A_2$ and $B_3$ sites, although the hopping that appears in Eq.~(\ref{eq:hamiltonian n-ABC}) is the strong interlayer coupling of the orbitals on the dimer $B_i$ and $A_{i+1}$ sites. The validity of the approximation is based on the increase in energy near the dimer atomic sites. For low Fermi energy, it therefore makes sense to take into account only the orbital wave functions near the other two atoms, i.e. the non-dimer sites.\cite{McCann2013, van2013multiband} The eigenenergies $E_{p,s}^{n}$ and the corresponding eigenstates $\Psi_{\vec{p},s}^{n}$ of the Hamiltonian (\ref{eq:hamiltonian n-ABC}) can be expressed as
\begin{equation}
E_{\vec{p},s}^{n}=s\frac{p^{n}}{\gamma},\label{eq:eigenernegy}
\end{equation}
and
\begin{equation}
\Psi_{\vec{p},s}^{n}=\frac{1}{\sqrt{2}}\left[\begin{array}{c}
1\\
se^{in\phi}
\end{array}\right],\label{eq:eigenstates}
\end{equation}
where $s=1$ ($s=-1$) is the electron conduction (hole valence) band index, $p=\hbar k$, $\gamma^{n-1}/v_{F}^{n} \rightarrow \gamma$ and $e^{i\phi}=\left(p_{x}+ip_{y}\right)/p$. This continuum approximation is valid in the low-energy and long-wavelength limits, and a small quantitative deviation of this approximation becomes more significant for large $k$ values as shown in Fig.~\ref{Fig: Geometries-1}(c) by comparing the energy spectrum obtained by the two-band continuum (black solid curves) and tight-binding (red dashed curves) models for mono ($n=1$), bi ($n=2$), tri ($n=3$) and tetra ($n=4$) layer graphene. Notice that for $n=1$, both multi-band [Eq.~(\ref{eq:hamiltonian-n layers})] and two-band [Eq.~(\ref{eq:hamiltonian n-ABC})] models give the same results, as already expected since each matrix element in the main diagonal in Eq.~(\ref{eq:hamiltonian-n layers}) represents a MLG Hamiltonian (see first left panel in Fig.~\ref{Fig: Geometries-1}(c)). This good agreement for the lowest two bands and near the Dirac cone has been widely reported and used in multilayer graphene works in the literature [for example, see Refs.~\citep{Duppen2013, van2013multiband,bala2012chiral, Partoens2006, PhysRevB.75.193402, Kumar2012, Prarokijjak2015,McCann2007}. Futhermore, similar works aiming the zitterbewegung investigation in multiband Hamiltonian with arbitrary matrix elements depending only on the momentum of the quasiparticle have been reported,\citep{PhysRevB.81.121417, PhysRevB.82.201405} showing the applicability and versatility of this kind of multiband-type model.

\subsection{Gaussian wave packet dynamics for $ABC$-NLG\label{subsec: wave function}}

Using the Green's function method, we obtained, inspired in the monolayer graphene case presented by Maksimova \textit{et al}.\citep{Maksimova2008} and Demikhvskii \textit{et al}.\citep{Demikhovskii2008}, a generalized expression to study ZBW in $ABC$-NLG.

According to Eqs.~(\ref{eq:eigenernegy}) and (\ref{eq:eigenstates}),
the time-dependent eigenfunctions of Hamiltonian (\ref{eq:hamiltonian n-ABC})
are given by 
\begin{equation}
\Phi_{p,s}\left(\vec{r},t\right)=\frac{1}{2\sqrt{2}\pi\hbar}\text{exp}\left(i\frac{\vec{p}\cdot\vec{r}}{\hbar}-i\frac{E_{\vec{p},s}^{n}t}{\hbar}\right)\left(\begin{array}{c}
1\\
se^{in\phi}
\end{array}\right).\label{eq:eigenfunctions}
\end{equation}

In order to calculate the time evolution of an arbitrary state, we use the Green's function method defined by the non-diagonal $2\times2$ matrix 
\begin{equation}
\mathbb{G}=\left(\begin{array}{cc}
G_{11} & G_{12}\\
G_{21} & G_{22}
\end{array}\right),
\end{equation}
where the matrix elements can be written as
\begin{equation}
G_{\mu v}\left(\vec{r},\vec{r}',t\right)=\sum_{s=\pm1}\int \Phi_{p,s,\mu}\left(\vec{r},t\right)\Phi_{p,s,v}^{\dagger}\left(\vec{r}',0\right)d\vec{p},\label{eq:greens functions}
\end{equation}
and $\mu,\nu=1,2$ are matrix indices, associated with the upper and lower components of $\Psi\left(\vec{r},t\right)$ that are related to the probability of finding the electron at the sublattices $A$ (upper) and $B$ (lower). The time-evolved electron wave function for $t>0$ can be obtained as
\begin{equation}
\Psi_{\mu}\left(\vec{r},t\right)=\int G_{\mu v}\left(\vec{r},\vec{r}',t\right)\psi_{v}\left(\vec{r},0\right)d\vec{r}'.\label{eq:arbitrary wave packet}
\end{equation}
Combining Eqs.~(\ref{eq:eigenfunctions}) and (\ref{eq:greens functions}),
we have that
\begin{widetext}
\begin{subequations}
\begin{equation}
G_{11}\left(\vec{r},\vec{r}',t\right)=G_{22}\left(\vec{r},\vec{r}',t\right)=\frac{1}{\left(2\pi\hbar\right)^{2}}\int\text{exp}\left[i\frac{\vec{p}\left(\vec{r}-\vec{r}'\right)}{\hbar}\right]\text{cos}\left(\frac{p^{n}t}{\gamma\hbar}\right)d\vec{p},\label{eq:G11 and G22}
\end{equation}
\begin{equation}
G_{12(-)}\left(\vec{r},\vec{r}',t\right)=G_{21(+)}\left(\vec{r},\vec{r}',t\right)=\frac{-i}{\left(2\pi\hbar\right)^{2}}\int e^{\mp in\phi}\text{exp}\left[i\frac{\vec{p}\left(\vec{r}-\vec{r}'\right)}{\hbar}\right]\text{sin}\left(\frac{p^{n}t}{\gamma\hbar}\right)d\vec{p}.\label{eq:G12 and G21}
\end{equation}
\end{subequations}
\end{widetext}
Note that $G_{12}\left(\vec{r},\vec{r}',t\right)$ differs from $G_{21}\left(\vec{r},\vec{r}',t\right)$
only by a negative sign in the term $e^{\mp in\phi}=\left(p_{x}\mp ip_{y}/p\right)^{n}$, as emphasized by the subscripts in Eq. (\ref{eq:G12 and G21}).

At $t=0$, we assume the wave function to be a circularly symmetrical 2D Gaussian wave packet with width $d$ and non-vanishing average momentum along $y$-direction, i.e. $p_{0y}=\hbar k_{0}^{y}$, such that
\begin{subequations}
\begin{equation}
\psi\left(\vec{r},0\right)=\frac{f\left(\vec{r}\right)}{\sqrt{\left|C_{1}\right|^{2}+\left|C_{2}\right|^{2}}}\begin{pmatrix}C_{1}\\
C_{2}
\end{pmatrix},\label{eq:initial wave}
\end{equation}
with
\begin{equation}
f\left(\vec{r}\right)=\frac{1}{d\sqrt{\pi}}\text{exp}\left[-\frac{r^{2}}{2d^{2}}+\frac{ip_{0y}y}{\hbar}\right].\label{eq:initial wave 2}
\end{equation}
\end{subequations}
Gaussian-like wave packets are commonly used in the ZBW analysis, since such oscillatory effect is not a stationary state but a dynamical phenomenon as well as it exhibits a minimal position-momentum uncertainty. They are essentially a combination of plane-waves, where the wave packet width represents a distribution of momenta and, consequently, of energy, and it is associated with e.g. the temperature of the system. Thus, by setting the initial state as Gaussian wave packet, this assumption covers most cases of practical interest, because any wave packet can be approximated by a superposition of a finite number of Gaussian states. Such a wave packet could be created by an ultra short laser pulse. This results in a wave packet with both positive and negative energies, since such a pulse has a very wide frequency spectrum \cite{Rusin2009, rusin2014zitterbewegung}.

Coefficients $C_{1}$ and $C_{2}$ determine the initial pseudospin polarization of the injected wave packet and are related to the two pseudospin components in Eq.~(\ref{eq:eigenstates}). Each component of the electron spinor wave function is then found as
\begin{equation}
\left(\hspace{-0.18cm}\begin{array}{c}
\Psi_{1}\left(\vec{r},t\right)\\
\Psi_{2}\left(\vec{r},t\right)
\end{array}\hspace{-0.18cm}\right)\hspace{-0.1cm}=\hspace{-0.1cm}\frac{1}{\sqrt{\left|C_{1}\right|^{2}\hspace{-0.1cm}+\hspace{-0.1cm}\left|C_{2}\right|^{2}}}\left(\hspace{-0.18cm}\begin{array}{c}
C_{1}\Phi_{1}\left(\vec{r},t\right)\hspace{-0.1cm}+\hspace{-0.1cm}C_{2}\Phi_{3}\left(\vec{r},t\right)\\
C_{1}\Phi_{2}\left(\vec{r},t\right)\hspace{-0.1cm}+\hspace{-0.1cm}C_{2}\Phi_{4}\left(\vec{r},t\right)
\end{array}\hspace{-0.18cm}\right),\label{eq:two components}
\end{equation}
\begin{widetext}
where
\begin{subequations}
\begin{equation}
\Phi_{1}\left(\vec{r},t\right)=\int G_{11}\left(\vec{r},\vec{r}',t\right)f\left(\overrightarrow{r}'\right)d\vec{r}'=\frac{de^{-\frac{\left(k_{0}^{y}d\right)^{2}}{2}}}{2\hbar^{2}\sqrt{\pi^{3}}}\int\text{exp}\left(i\frac{\vec{p}\cdot\vec{r}}{\hbar}-\frac{p^{2}d^{2}}{2\hbar^{2}}+\frac{p_{y'}k_{0}^{y}d^{2}}{\hbar}\right)\text{cos}\left(\frac{p^{n}t}{\gamma\hbar}\right)d\vec{p},\label{eq:PHI1 PHI4}
\end{equation}
\begin{equation}
\Phi_{3_{-}\left(2_{+}\right)}\left(\vec{r},t\right)\hspace{-0.5mm}=\hspace{-0.5mm}\int G_{12\left(21\right)}\left(\vec{r},\vec{r}',t\right)\hspace{-0.5mm}f\hspace{-0.5mm}\left(\overrightarrow{r}'\right)d\vec{r}'\hspace{-0.5mm}=\hspace{-0.5mm}\frac{-ide^{-\frac{\left(k_{0}^{y}d\right)^{2}}{2}}}{2\hbar^{2}\sqrt{\pi^{3}}}\hspace{-0.5mm}\int \hspace{-0.5mm}e^{\mp in\phi}\text{exp}\hspace{-0.2mm}\left(\hspace{-0.5mm}i\frac{\vec{p}\cdot\vec{r}}{\hbar}\hspace{-0.5mm}-\hspace{-0.5mm}\frac{p^{2}d^{2}}{2\hbar^{2}}\hspace{-0.5mm}+\hspace{-0.5mm}\frac{p_{y'}k_{0}^{y}d^{2}}{\hbar}\right)\hspace{-0.5mm}\text{sin}\hspace{-0.5mm}\left(\frac{p^{n}t}{\gamma\hbar}\right)\hspace{-0.5mm}d\vec{p},\label{eq:PHI2 PHI3}
\end{equation}
\end{subequations}
\end{widetext}and $\Phi_{1}\left(\vec{r},t\right)=\Phi_{4}\left(\vec{r},t\right)$
according to Eq.~(\ref{eq:G11 and G22}).
The subscript $-$ ($+$) for $\Phi_{3}$ ($\Phi_{2}$) in Eq.~(\ref{eq:PHI2 PHI3}) refers to the sign of the argument in $e^{-in\phi}$ ($e^{+in\phi}$).

Using cylindrical coordinates in Eqs.~(\ref{eq:PHI1 PHI4}) and (\ref{eq:PHI2 PHI3})
and integrating over the angular variable (see Appendix for more details), we obtain
\begin{widetext}
\begin{subequations}
\begin{equation}
\Phi_{1}\left(\vec{r},t\right)=\frac{e^{-a^{2}/2}}{d\sqrt{\pi}}\int_{0}^{\infty}e^{-\frac{q^{2}}{2}}\text{cos}\left(q^{n}t'\right)J_{0}\left(q\sqrt{r^{2}-a^{2}-2iay}\right)qdq,\label{eq:PHI1 PHI4-1}
\end{equation}
\begin{equation}
\Phi_{3_{+}\left(2_{-}\right)}\left(\vec{r},t\right)=\frac{-ie^{-a^{2}/2}}{d\sqrt{\pi}}\left[\frac{ix'\pm y\mp ia}{\sqrt{r^{2}-a^{2}-2iay}}\right]^{n}\int_{0}^{\infty}e^{-\frac{q^{2}}{2}}\text{sin}\left(q^{n}t'\right)J_{n}\left(q\sqrt{r^{2}-a^{2}-2iay}\right)qdq,\label{eq:PHI2 PHI3-1}
\end{equation}
\end{subequations}
\end{widetext}
where $J_{0}\left(z\right)$ and $J_{n}\left(z\right)$ are Bessel
functions of the zeroth and $n$-th order. For the sake of simplicity, we introduced in Eqs.~(\ref{eq:PHI1 PHI4-1}) and (\ref{eq:PHI2 PHI3-1}) the dimensionless parameter $a=k_{0}^{y}d$ and considered the time in units of $d/v_{F}$. 

Once $\Psi_{1}\left(\vec{r},t\right)$ and $\Psi_{2}\left(\vec{r},t\right)$
are known, the time-dependent expectation value of the position operator can be more calculated as
\begin{equation}
\left\langle \vec{r}\left(t\right)\right\rangle =\sum_{j=1}^{2}\int\Psi_{j}^{*}\left(\vec{p},t\right)\left[i\hbar\frac{d}{d\vec{p}}\right]\Psi_{j}\left(\vec{p},t\right)d\vec{p},
\label{eq:position operator}
\end{equation}
with $\Psi$ in momentum representation, that can be easily inferred from Eqs. (\ref{eq:PHI1 PHI4}) and (\ref{eq:PHI2 PHI3}). From Eq.~(\ref{eq:position operator})
we investigate the ZBW phenomenon by an analytical calculation
of the time-dependent expectation value of the position $\left\langle \vec{r}\left(t\right)\right\rangle =(\left\langle x\left(t\right)\right\rangle ,\left\langle y\left(t\right)\right\rangle )$
of the center of the wave packet for different initial electron amplitudes of sublattices $A$ and $B$, by taking different values for $C_{1}$ and $C_{2}$ in Eq.~(\ref{eq:two components}),
as will be discussed in Sec.\ref{sec: Results}.

\subsection{SOT for $ABC$-NLG within Dirac model\label{subsec: Split-operator-technique}}

The analytical method developed here so far, despite being exact, is not flexible enough to allow the study of wave packet propagation in $ABC$-NLG in the presence of e.g. external potentials and applied electric or magnetic fields. We, thus, propose here a semi-analytical method, namely, the SOT, \citep{Chaves2010,Pereira2010,Chaves2015,Rakhimov2011,Costa2015,da2012wave, cavalcante2016all, chaves2015energy, da2017valley, abdullah2019electron} which consists in splitting the time-evolution operator $\exp\left[-\frac{i}{\hbar}\mathcal{H}\Delta t\right]$ into different terms involving the potential $\mathcal{V}$, in real space, and the kinetic energy $\mathcal{H}_k$, in reciprocal space:
\begin{equation}
e^{\left[-\frac{i}{\hbar}\mathcal{H}\Delta t\right]} = e^{\left[-\frac{i}{2\hbar}\mathcal{V}\Delta t\right]}e^{\left[-\frac{i}{\hbar}\mathcal{H}_k\Delta t\right]}e^{\left[-\frac{i}{2\hbar}\mathcal{V}\Delta t\right]} + O(\Delta t^3). \label{eq:SOT}
\end{equation}
The error of order $\Delta t^3$ comes from the non-commutativity between potential and kinetic energy operators, and can be made small by assuming small time steps. 

As an example, let's consider the Dirac Hamiltonian for MLG\citep{Neto2009} in the absence of external potentials ($V = 0$), i.e.
\begin{equation}
H_{MLG}=v_{F}\vec{\sigma}\cdot\vec{p}.\label{eq:monolayer}
\end{equation}
The time evolution operator for this case can be written as 
\begin{equation}
\text{exp}\hspace{-0.7mm}\left[\hspace{-0.7mm}-\frac{i}{\hbar}\mathcal{H}_{MLG}\Delta t\right]\hspace{-0.7mm}=\hspace{-0.7mm}\text{exp}\hspace{-0.7mm}\left[\hspace{-0.7mm}-\frac{iv_{F}}{\hbar}\hspace{-0.7mm}\left(\vec{p}\cdot\vec{\sigma}\right)\hspace{-0.7mm}\Delta t\hspace{-0.5mm}\right]\hspace{-0.7mm}=\hspace{-0.7mm}\text{exp}\hspace{-0.7mm}\left[-i\vec{S}\cdot\vec{\sigma}\right],\label{eq: time-evolution operator}
\end{equation}
where $\vec{S}=\Delta tv_{F}\vec{p}/\hbar$ and its magnitude is $S=\Delta tv_{F}\sqrt{k_{x}^{2}+k_{y}^{2}}$. Using the properties of the Pauli matrices, it is possible to rewrite Eq.~(\ref{eq: time-evolution operator})
as a sum of two matrices, such as
\begin{equation}\label{eq:expansionPauli}
\text{exp}\left[-i\vec{S}\cdot\vec{\sigma}\right]=\text{cos}\left(S\right)\mathbb{1}-i\frac{\text{sin}\left(S\right)}{S}\left(\vec{S}\cdot\vec{\sigma}\right)=\mathbb{M},
\end{equation}
where $\mathbb{1}$ denotes the $2\times 2$ unit matrix. This is an exact representation of the time evolution operator, including all the terms of the expansion of the exponential. 

The generalized Hamiltonian $H_n$ for $ABC$-NLG, Eq.~(\ref{eq:hamiltonian n-ABC}), can be re-written in terms of Pauli matrices for any number of layers $n$, therefore, Eq.~(\ref{eq:expansionPauli}) always hold, as long as the vector $\vec{S}$ one adapts accordingly, which can be done with straightforward algebra. For instance, for BLG one can re-write $\vec{S}$ as 
\begin{equation}
\vec{S}=\hbar v_{F}^{2}\Delta t\gamma^{-1}\left(k_{x}^{2}-k_{y}^{2},2k_{x}k_{y},0\right),    
\end{equation}
whereas for TLG, one obtains
\begin{equation}
\vec{S}=\hbar^2 v_{F}^{3}\Delta t\gamma^{-2}\left(k_{x}^{3}-3k_{y}^{2}k_{x},3k_{x}^2k_{y}-k_y^3,0\right).    
\end{equation}

The propagated wave function $\Psi = [\Psi_1 ~\Psi_2]^T$ at a time step $t+\Delta t$ is given by
\begin{equation}
\Psi\left(\vec{r},t+\Delta t\right)=e^{-i{H}_{n}\Delta t/\hbar}\Psi\left(\vec{r},t\right)=\mathbb{M}\Psi\left(\vec{r},t\right).\label{eq:SP - wave function}
\end{equation}
Note that $\mathbb{M}$ depends on the wave vectors $k_{x}$ and $k_{y}$, therefore, the matrix multiplication with a general initial wave packet is conveniently computed numerically in reciprocal space by performing a Fourier transform of the wave function, reason why this method is thus seen as a semi-analytical procedure. Because the solution of Eq.~(\ref{eq:SP - wave function}) is exact, it should provide the same results as the Green's function method described in Sec.~\ref{subsec: wave function} for free wave packets in NLG. We verified, as will be discussed latter in Sec.~\ref{sec: Results}, that we obtain numerical \textit{perfect} agreement between results obtained by the SOT and the Green's function formalism. A clear advantage of the SOT is that it provides a way to study the wave packet dynamics in NLG within the continuum model in the presence of arbitrary external potential profiles\cite{Chaves2010,Pereira2010,Chaves2015,Rakhimov2011,Costa2015,da2012wave, cavalcante2016all, chaves2015energy, da2017valley, abdullah2019electron}, simply by performing matrix multiplications with the potential exponential terms, as shown in Eq.~(\ref{eq:SOT}).

\subsection{SOT for $ABC$-NLG within the tight-binding model}

Despite having the advantage of being semi-analytical, numerically exact, and suitable for large graphene samples, the methods developed here so far are not able to capture the microscopic features of NLG, such as rough edges and lattice defects. For that, one needs to invoke theories that deal with the 2D material on the microscopic level, such as the density functional theory and the tight-binding model. Nevertheless, for the later, the SOT has been already developed for MLG\cite{Chaves2010, Chaves2015} and BLG\cite{Costa2015} cases. Details of this procedure and the method proposed in Ref.~[\onlinecite{Costa2015}] can be easily adapted for any number of layers, but such fully numerical microscopic approach is beyond the scope of the present work. Although not shown in this paper, the time evolution of wave packets and trajectories obtained here for all cases of wave packet pseudospinor are verified to agree well with those one based on the tight-binding SOT for low-energy wave packets in MLG\cite{Chaves2010,Pereira2010,Chaves2015,Rakhimov2011,da2012wave, cavalcante2016all, chaves2015energy, da2017valley} and BLG\cite{Costa2015,abdullah2019electron}, thus additionally validating our results.  

\section{ZITTERBEWEGUNG OF GAUSSIAN WAVE PACKET FOR DIFFERENT PSEUDOSPIN POLARIZATIONS\label{sec: Results}}

\subsection{Predictions from the Heisenberg equation\label{subsec:Heisenberg-Picture}}

Different kinds of initial pseudospin polarization of the wave packet will be considered in this work. It is thus important to be able to predict beforehand the qualitative behavior of the propagating wave packet in each case. In order to do so, we introduce a method based on calculations of expectation values of wave packets by using the Heisenberg equation.

We use the subtlety of Heisenberg representation to predict which initial settings of pseudospin $\left(C_{1}\  C_{2}\right)^{T}$ result in non-zero averages of the electron coordinates $\left\langle x\left(t\right)\right\rangle $ and $\left\langle y\left(t\right)\right\rangle $. The velocity vector is defined as
\begin{equation}
\left\langle \vec{v}\left(t\right)\right\rangle =\frac{d\vec{r}}{dt}=\frac{1}{i\hbar}\left[\vec{r},H\right]=v_{F}\vec{\sigma},\label{eq:heisenberg picture}
\end{equation}
where $\vec{v}=\left(v_{x},v_{y}\right)$ and $\vec{r}=\left(x,y\right)$ are the velocity and the position vectors, respectively. 

Without loss of generality, as an example, let's consider the MLG Hamiltonian [Eq.~(\ref{eq:monolayer})] and shall analyse a wave packet propagating in the $x$-direction in order to verify whether $\left\langle x\left(t\right)\right\rangle $ is a constant of motion. Therefore, from Eqs.~(\ref{eq:monolayer})  and (\ref{eq:heisenberg picture}), one obtains
\begin{equation}
\frac{d\left\langle x\left(t\right)\right\rangle }{dt}=\frac{1}{i\hbar}\left\langle \left[x,H_{MLG}\right]\right\rangle =v_{F}\left\langle \sigma_{x}\right\rangle .\label{eq:Heisemberg - <x>}
\end{equation}
On the other hand,
\begin{equation}
\frac{d\left\langle \sigma_{x}\right\rangle }{dt}=\frac{1}{i\hbar}\left[\sigma_{x},H_{MLG}\right]=\frac{2v_{F}p_y}{\hbar}\left\langle \sigma_{z}\right\rangle .\label{eq:heisenberg picture-1}
\end{equation}
Thus, from Eqs.~(\ref{eq:Heisemberg - <x>}) and (\ref{eq:heisenberg picture-1}), we conclude that, if the initial pseudospin is oriented along the $z$ direction, i.e., $\left\langle \sigma_{z}\right\rangle \neq0$, and $p_y \neq 0$, $\left\langle x\left(t\right)\right\rangle $ is not a constant of motion and it is expected that $\left\langle x\left(t\right)\right\rangle $ will exhibit ZBW. This choice is represented by the initial pseudospinor $\left(C_{1}\ C_{2}\right)^{T}=\left(1\ 0\right)^{T}$. The same idea is straightforwardly generalized to any number of layers. Table \ref{Tab: Heisemberg Picture} shows the results for MLG, BLG and TLG for other initial pseudospin configurations, which are the three cases developed in detail in the next sections.

\begin{table}[H]
\centering{}\caption{Expectation value of the position $(x,y)$ of the injected wave packet obtained from the Heisenberg picture for different $C_{1}$ and $C_{2}$ values that determine the initial polarization of the pseudospin. The ($\neq$) $=$ symbols indicate expectation values that are (non-)zero.}
 
\begin{tabular}{ccccccccc}
\hline\hline

& &  \multicolumn{3}{c}{$\left\langle x\left(t\right)\right\rangle $} && \multicolumn{3}{c}{$\left\langle y\left(t\right)\right\rangle$}\\ 

\hline
$\left(C_{1}\ C_{2}\right)^{T}$   && $\left(1\ 0\right)^{T}$     & $\left(1\ 1\right)^{T}$ &  $\left(1\ i\right)^{T}$   && $\left(1\ 0\right)^{T}$ & $\left(1\ 1\right)^{T}$ & $\left(1\ i\right)^{T}$\\
\hline
Monolayer &  &  $\neq$ &    $\neq$ & $=$ &  & $=$ & $=$ & $\neq$ \\
Bilayer   &  &  $\neq$ &    $=$ & $\neq$ &  & $=$ & $\neq$ & $=$ \\
Trilayer  &  &  $\neq$ &    $=$ & $\neq$ &  & $=$ & $\neq$ & $=$ \\

\hline\hline
\end{tabular}\label{Tab: Heisemberg Picture}
\end{table}

\subsection{ZBW in MLG\label{subsec:MONOLAYER-GRAPHENE-CASE}}

Note that Eqs.~(\ref{eq:PHI1 PHI4}) and (\ref{eq:PHI2 PHI3}) were generally obtained for NLG. Thus, one just needs to use $n=1$ in these equations and replace them into Eq.~(\ref{eq:two components}) in order to obtain the wave function for MLG. Once the wave function is obtained, the expectation value of the position of its center of mass is calculated using Eq.~(\ref{eq:position operator}). Let us first revisit the problem of ZBW in MLG as a particular case of the method developed here.

\begin{figure}[!t]
\centering \includegraphics[width=1\columnwidth]{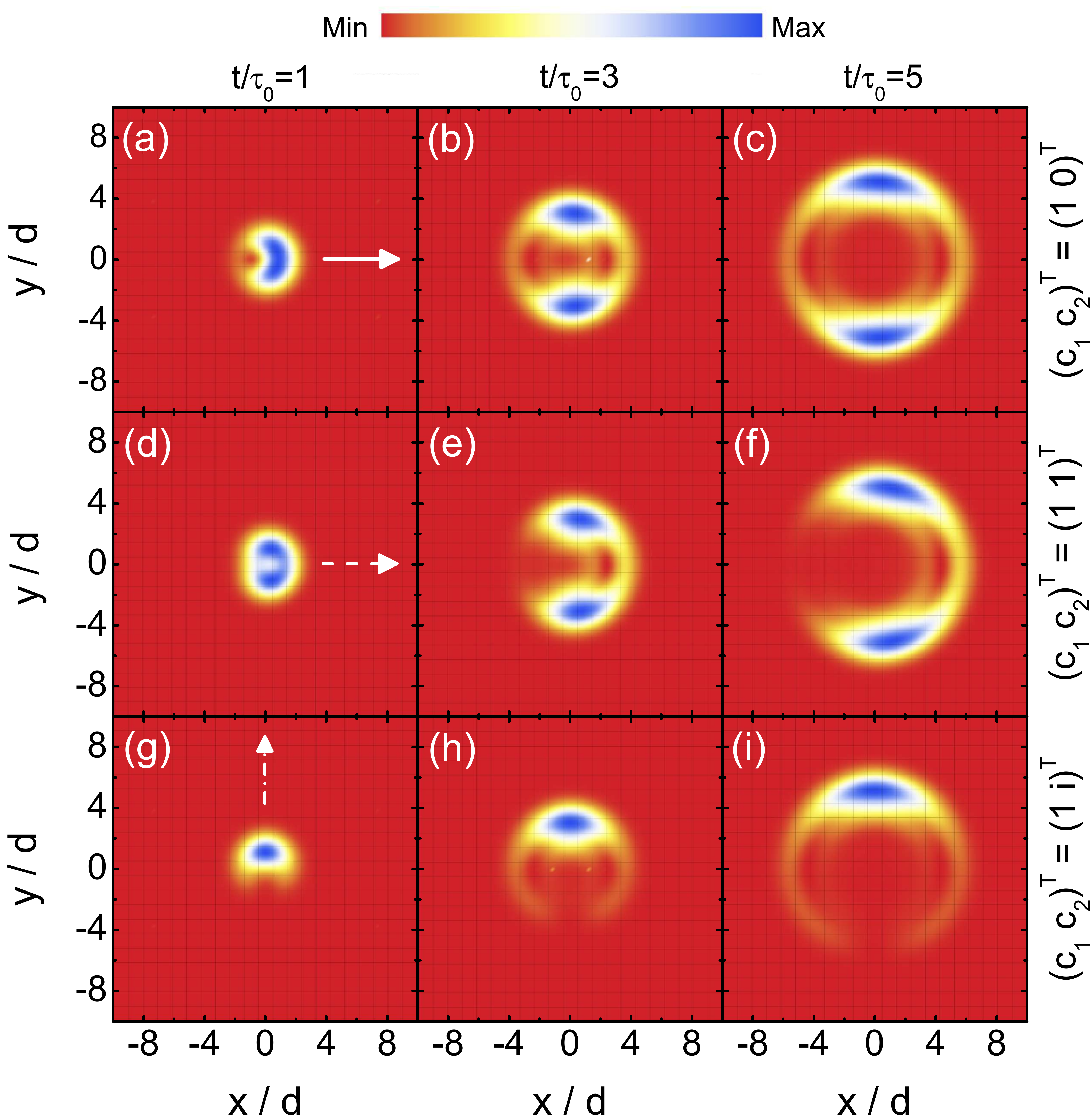}\caption{(Color online) Evolution (in units of $d/v_{F}$) of electronic probability density $\rho\left(\vec{r},t\right)=\left|\Psi_{1}\left(\vec{r},t\right)\right|^{2}+\left|\Psi_{2}\left(\vec{r},t\right)\right|^{2}$ for MLG with (a)-(c) $\left(C_{1}\ C_{2}\right)^{T}=\left(1\ 0\right)^{T}$,  (d)-(f) $\left(C_{1}\ C_{2}\right)^{T}=\left(1\ 1\right)^{T}$, (g)-(i) $\left(C_{1}\ C_{2}\right)^{T}=\left(1\ i\right)^{T}$, for $a=k_{0}^{y}d=1.2$ ($d=2\ \text{nm}$ and $k_{0}^{y}=0.6\ \text{nm}^{-1}$) and $t/\tau_{0}=1$, 3 and 5. The white arrows indicate the direction of propagation of the wave packet.}
\label{Fig Mono: Densidade - Monolayer}
\end{figure}

\subsubsection{$C_{1}=1$ and $C_{2}=0$\label{Subsub: Mono C1=00003D1 e C2=00003D0}}

We first consider the simple case when the lower component of the initial wave function (\ref{eq:initial wave}) is equal to zero, i.e. taking $C_{1}=1$ and $C_{2}=0$ in Eq.~(\ref{eq:two components}). It corresponds to the case in which the electron probability is initially located only at sites of the sublattice $A$ and pseudospin is polarized perpendicularly to the $xy$-plane, i.e., $\left\langle \sigma_{z}\right\rangle =1$ and $\left\langle \sigma_{x}\right\rangle =\left\langle \sigma_{y}\right\rangle =0$.

\begin{figure}[t]
\centering \includegraphics[width=1\columnwidth]{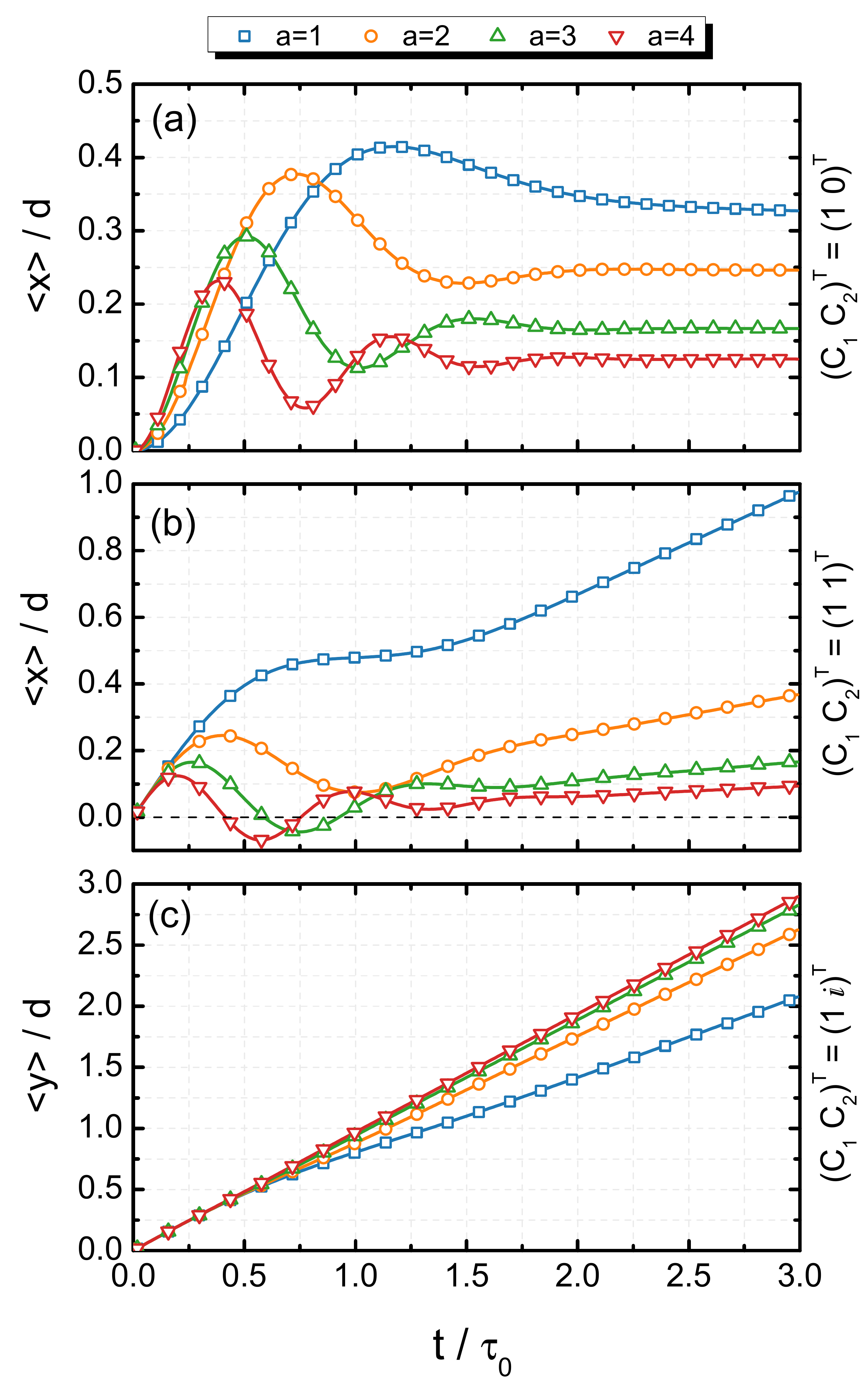}\caption{(Color online) (a) Expectation value $\left\langle x\left(t\right)\right\rangle$ of the Gaussian wave packet center-of-mass as a function of time ($\tau_{0}=d/v_{F}$) for MLG with pseudospin polarization (a) $\left(C_{1}\ C_{2}\right)^{T}=\left(1\ 0\right)^{T}$, (b) $\left(C_{1}\ C_{2}\right)^{T}=\left(1\ 1\right)^{T}$ and (c) $\left(C_{1}\ C_{2}\right)^{T}=\left(1\ i\right)^{T}$, for different values of $a=k_{0}^{y}d$. The results are obtained for a fixed value of wave packet width $d=100\ \lyxmathsym{\protect\AA}$ and different initial $y$-momentum: $k_{0}^{y}=1\cdot10^{-2}\ \lyxmathsym{\protect\AA}^{-1}$ (blue); $k_{0}^{y}=2\cdot10^{-2}\ \lyxmathsym{\protect\AA}^{-1}$ (orange); $k_{0}^{y}=3\cdot10^{-2}\ \lyxmathsym{\protect\AA}^{-1}$ (green) and $k_{0}^{y}=4\cdot10^{-2}\ \lyxmathsym{\protect\AA}^{-1}$ (red). The solid curves (symbols) correspond to the results obtained by the Green's function (SOT) method.}
\label{Fig Mono: Expectation values}
\end{figure}

According to Eq.~(\ref{eq:two components}), the wave function for
$t>0$ has the form:
\begin{equation}
\left(\hspace{-0.15cm}\begin{array}{c}
\Psi_{1}\left(\vec{r},t\right)\\
\Psi_{2}\left(\vec{r},t\right)
\end{array}\hspace{-0.15cm}\right)=\left(\hspace{-0.15cm}\begin{array}{c}
\Phi_{1}\left(\vec{r},t\right)\\
\Phi_{2}\left(\vec{r},t\right)
\end{array}\hspace{-0.15cm}\right),\label{eq: Mono - Wave function (1,0)}
\end{equation}
where $\Phi_{1,2}\left(\vec{r},t\right)$ are defined by Eqs.~(\ref{eq:PHI1 PHI4}) and (\ref{eq:PHI2 PHI3}), respectively, with $n=1$. To illustrate the evolution of the electron probability density we show $\rho\left(\vec{r},t\right)=\left|\Psi_{1}\left(\vec{r},t\right)\right|^{2}+\left|\Psi_{2}\left(\vec{r},t\right)\right|^{2}$ in Fig.~\ref{Fig Mono: Densidade - Monolayer}(a)-(c) for $p_{0y}=\hbar k^y_{0}\neq0$. Inset in Fig.~\ref{Fig Mono: Densidade - Monolayer}(a) shows the projection of the 2D Gaussian wave packet centered in the $xy$-plane at $t=0$. As time elapses, the wave packet splits into two parts moving along the $y-$axis with opposite speeds, Figs.~\ref{Fig Mono: Densidade - Monolayer}(a)-\ref{Fig Mono: Densidade - Monolayer}(c). The probability density is symmetric (asymmetric) with respect to $y$ ($x$), i.e., $\rho\left(x,y,t\right)=\rho\left(x,-y,t\right)$ ($\rho\left(x,y,t\right)\neq\rho\left(-x,y,t\right)$). Thus, the center of the wave packet oscillates (ZBW) only along the $x$-direction. For long enough time, the width of the wave packet increases due to the effect of dispersion~\footnote{In fact, this is true for all other cases of pseudo-spin and number of graphene layers.} as for the case of a free particle. This is unexpected, since the Dirac spectrum of low-energy electrons in graphene suggests a dispersionless wave function, thus the observed dispersion is a direct effect of the ZBW, as pointed out also in previous studies. \cite{Maksimova2008,Zawadzki2011,Zawadzki2010}

The expectation value of the position operator were obtained by inserting Eq.~(\ref{eq: Mono - Wave function (1,0)}) into Eq.~(\ref{eq:position operator}), which leads
\begin{equation}
\left\langle x\left(t\right)\right\rangle \hspace{-0.75mm}=\hspace{-0.75mm}d\left[\hspace{-1mm}\frac{1-e^{-a^{2}}}{2a}\hspace{-0.2mm}-\hspace{-0.2mm}e^{-a^{2}}\hspace{-2mm}\int_{0}^{\infty}\hspace{-2mm}e^{-q^{2}}\text{\text{cos}}\left(2qt'\right)\hspace{-0.75mm}I_{1}\hspace{-0.75mm}\left(2aq\right)dq\hspace{-0.1mm}\right]
\label{eq: <x(t)> - (1,0)}
\end{equation}
and $\left\langle y\left(t\right)\right\rangle =0$, where $I_{1}\left(z\right)$ is the modified Bessel function of the first order. These results are in accordance with Table \ref{Tab: Heisemberg Picture}, only obtained from the Heisenberg picture, and depends on the parameter $a=k_{0}^{y}d$.

The average position of the $x$-coordinate as a function of time, given by Eq.~(\ref{eq: <x(t)> - (1,0)}), is shown in Fig.~\ref{Fig Mono: Expectation values}(a) assuming various values of the parameter $a=k_{0}^{y}d$. For comparison, results obtained by the SOT based on the Dirac model are shown with symbols, presenting a good agreement with the analytical ones. From Fig.~\ref{Fig Mono: Expectation values}(a), the oscillations disappear after $t/\tau_0\approx 2.5$ and $\left\langle x\left(t\right)\right\rangle $ converges to a specific value given by the first term of Eq.~(\ref{eq: <x(t)> - (1,0)}). For example, for $a=4$, the first term in Eq.~(\ref{eq: <x(t)> - (1,0)}) is equal to $0.125$ (in units of $d$), corresponding to the converged value of the red curve in Fig.~\ref{Fig Mono: Expectation values}. This demonstrates that the ZBW is not permanent, but a transient feature, as discussed also in Refs.~{[}\onlinecite{Lurie1970,Zawadzki2005}{]},
and it is due to the time-dependence of the second term in Eq.~(\ref{eq: <x(t)> - (1,0)}). It can be noticed also in Fig.~\ref{Fig Mono: Expectation values}(a), that more oscillations occur, but with smaller amplitudes, as $a$ increases. Consequently, the velocity $v_{x}=d\left\langle x\left(t\right)\right\rangle /dt$ oscillates with shorter period and smaller amplitude as $a$ increases.
Notice that $\left\langle \vec{r}\left(t\right)\right\rangle $, obtained here as a particular case of Eq.~(\ref{eq:two components}), coincide with corresponding formulas reported in Ref.~{[}\onlinecite{Maksimova2008}{]}.

\subsubsection{$C_{1}=1$ and $C_{2}=1$ \label{Subsub: Mono C1=00003D1 e C2=00003D1}}

For $\left(C_{1}\ C_{2}\right)^{T}=\left(1\ 1\right)^{T}$, the initial pseudospin lies along the $x-$axis with the wave function equally distributed on sublattices $A$ and $B$. From Eq.~(\ref{eq:two components}), one has
\begin{equation}
\left(\hspace{-0.15cm}\begin{array}{c}
\Psi_{1}\left(\vec{r},t\right)\\
\Psi_{2}\left(\vec{r},t\right)
\end{array}\hspace{-0.15cm}\right)=\frac{1}{\sqrt{2}}\left(\hspace{-0.15cm}\begin{array}{c}
\Phi_{1}\left(\vec{r},t\right)+\Phi_{3}\left(\vec{r},t\right)\\
\Phi_{1}\left(\vec{r},t\right)+\Phi_{2}\left(\vec{r},t\right)
\end{array}\hspace{-0.15cm}\right),\label{eq: Mono - Wave function (1,1)}
\end{equation}
with $\Phi_{1,2,3}\left(\vec{r},t\right)$ given by Eqs.~(\ref{eq:PHI1 PHI4-1}) and (\ref{eq:PHI2 PHI3-1}), respectively. It is important to point up that an initial wave packet in which the electron probability density occupies equally all sublattices is more realistic experimentally, as an expected configuration when one creates wave packets by illuminating samples with short laser pulses and also because for an infinite system the initial wave function should describe electronic bulk states spread over all sites around the center point of the Gaussian distribution.\cite{Cunha2019,Rusin2009,rusin2014zitterbewegung} The time-evolved electron probability densities for $\left(1\ 1\right)^{T}$ case are depicted in Fig.~\ref{Fig Mono:  Densidade - Monolayer}(c)-(e). For $t>0$, the shape of the full electron density $\rho\left(\vec{r},t\right)$ changes, see Figs.~\ref{Fig Mono:  Densidade - Monolayer}(c)-(e), splitting into two parts that move along the $y-$axis in opposite direction. As in the previous case, $\rho\left(\vec{r},t\right)$ is not mirror symmetric with respect to $x=0$ axis and the wave packet travels asymmetrically to the positive $x$-direction. Consequently, the motion of the center of the Gaussian wave packet oscillates (ZBW) only along this direction. This is illustrated by two maxima of the electron density spread along the $x$-direction.

By substituting Eq.~(\ref{eq: Mono - Wave function (1,1)}) into Eq.~(\ref{eq:position operator}), we obtain the time-dependent expectation value of the wave packet position
\begin{multline}
\left\langle x\left(t\right)\right\rangle =d\left(\frac{1-e^{-a^{2}}}{2a^{2}}\right)t\\
+\frac{de^{-a^{2}}}{2a}\int_{0}^{\infty}e^{-q^{2}}\text{sin}\left(2qt'\right)\left[\frac{d}{dq}I_{1}\left(2aq\right)\right]dq,\label{eq: Monolayerr <x(t)> - (1,1)}
\end{multline}
$\left\langle y\left(t\right)\right\rangle =0$.

Figure~\ref{Fig Mono: Expectation values}(b) presents $\left\langle x\left(t\right)\right\rangle $, given by Eq.~(\ref{eq: Monolayerr <x(t)> - (1,1)}), for different values of the parameter $a$ and demonstrates that: (i) the higher the value of $a$, the smaller the amplitude of the ZBW, the period of oscillations and the velocity $v_{x}$ of the center of the wave packet; and (ii) ZBW is transient. Results from SOT within the Dirac model are shown with symbols, and an excellent agreement with the analytical results (solid curves) validates our method. For small values of the wave packet initial momentum $k_{0}^{y}$, i.e. small values of $a=k_{0}^{y}d$, and after ZBW vanishes, one observes that $\left\langle x\left(t\right)\right\rangle $ increases linearly with time, as a consequence of the linear time-dependence on the first term of Eq.~(\ref{eq: Monolayerr <x(t)> - (1,1)}) that dominates after a while. However, as $a$ (or equivalently $k_0^y$) increases, the second integral term in Eq.~(\ref{eq: Monolayerr <x(t)> - (1,1)}) becomes the dominant one. 

\subsubsection{$C_{1}=1$ and $C_{2}=i$ \label{Subsub: Mono C1=00003D1 e C2=00003Di}}

In this last case, the initial pseudospin polarization $\left(C_{1}\ C_{2}\right)^{T}=\left(1\ i\right)^{T}$ is oriented along the same direction ($y$) as the plane wave momentum $p_{0y}$ in Eq.~(\ref{eq:initial wave 2}).
From Eq.~(\ref{eq:two components}), the wave function is given by
\begin{equation}
\left(\hspace{-0.15cm}\begin{array}{c}
\Psi_{1}\left(\vec{r},t\right)\\
\Psi_{2}\left(\vec{r},t\right)
\end{array}\hspace{-0.15cm}\right)=\frac{1}{\sqrt{2}}\left(\hspace{-0.15cm}\begin{array}{c}
\Phi_{1}\left(\vec{r},t\right)+i\Phi_{3}\left(\vec{r},t\right)\\
i\Phi_{1}\left(\vec{r},t\right)+\Phi_{2}\left(\vec{r},t\right)
\end{array}\hspace{-0.15cm}\right).\label{eq: Mono - Wave function (1,i)}
\end{equation}

Figures \ref{Fig Mono:  Densidade - Monolayer}(g)-(i) present snapshots of the propagated Gaussian wave packet for different time values. Unlike the two previous cases, discussed in  Secs.~\ref{Subsub: Mono C1=00003D1 e C2=00003D0} and \ref{Subsub: Mono C1=00003D1 e C2=00003D1}, the wave packet now moves along the $y-$axis, i.e. the wave packet travels along the same direction as the pseudospin and average momentum $p_{0y}$ orientation, and does not split into two parts for $t>0$. The electron probability density obeys the following symmetry (asymmetry) for $t>0$: $\rho\left(x,y,t\right)=\rho\left(-x,y,t\right)$ ($\rho\left(x,y,t\right)\neq\rho\left(x,-y,t\right)$).

Inserting Eq.~(\ref{eq: Mono - Wave function (1,i)}) into Eq.~(\ref{eq:position operator}), it is easy to show that the expectation values of the $x$ and $y$ coordinates are, respectively: $\left\langle x\left(t\right)\right\rangle =0$ and
\begin{multline}
\left\langle y\left(t\right)\right\rangle =d\left(1-\frac{1}{2a^{2}}+\frac{e^{-a^{2}}}{2a^{2}}\right)t\\
+\frac{de^{-a^{2}}}{2a}\int_{0}^{\infty}e^{-q^{2}}\text{sin}\left(2qt\right)\frac{I_{1}\left(2aq\right)}{q}dq.\label{eq: Monolayerr <y(t)> - (1,i)}
\end{multline}
Figure~\ref{Fig Mono: Expectation values}(c) compares the analytical results (solid curves) obtained by performing a numerical integration of Eq.~(\ref{eq: Monolayerr <y(t)> - (1,i)}), with those computed via SOT within the Dirac model (symbols). As can be seen from Fig.~\ref{Fig Mono: Expectation values}(c), the ZBW is almost absent and $\left\langle y\left(t\right)\right\rangle /d$ exhibits a linear time-dependence, which becomes more significant as the wave packet width $a$ increases, without significant oscillations. That is, $\left\langle y\left(t\right)\right\rangle /d\approx t$ for large $a$. According to Eq.~(\ref{eq: Monolayerr <y(t)> - (1,i)}), as $a$ increases, the second term (that causes oscillations), as well as the other two terms of the first expression which possess $a$ parameter in their denominators, become small. Therefore, only the linear term $t$ will dominate.

Our investigations reveal that the choice of the initial pseudospin polarization given by $(1\ i)^T$ is the best way, among the cases studied here, to avoid ZBW in MLG systems, as reported in Refs.~[\onlinecite{Chaves2010,Pereira2010,Chaves2015,Rakhimov2011,da2012wave, cavalcante2016all, chaves2015energy, da2017valley}]. Which is due to the fact that for this choice of pseudospin, the motion in the $y-$direction is perfectly vertical during the whole propagation (see Eq.~(\ref{eq: Monolayerr <y(t)> - (1,i)})), being the least affected by ZBW phenomena, specially moving straight without to much dispersion as larger is the initial Gaussian wave vector.

\subsection{ZBW in BLG\label{subsec:BILAYER-GRAPHENE-CASE}}

Owing to the distinct electronic and transport properties for graphene samples with different number of stacked layers, we also analyze the influence of the number of layers on the wave packet propagation with different pseudospin polarization, as well as we will verify which are the main ZBW features observed in NLG. We consider in the current section the BLG case and TLG will be investigated in next Sec.~\ref{subsec:trilayer-GRAPHENE-CASE}. 

The wave function is obtained by taking $n=2$ in Eqs.~(\ref{eq:PHI1 PHI4}) and (\ref{eq:PHI2 PHI3}) and combining them with Eq.~(\ref{eq:two components}). Once the wave function evolves in time, its $(x,y)$ position expectation values are calculated using Eq.~(\ref{eq:position operator}).

\subsubsection{$C_{1}=1$ and $C_{2}=0$\label{subsec:BLG - (1,0)}}

\begin{figure}[t]
\centering \includegraphics[width=1\columnwidth]{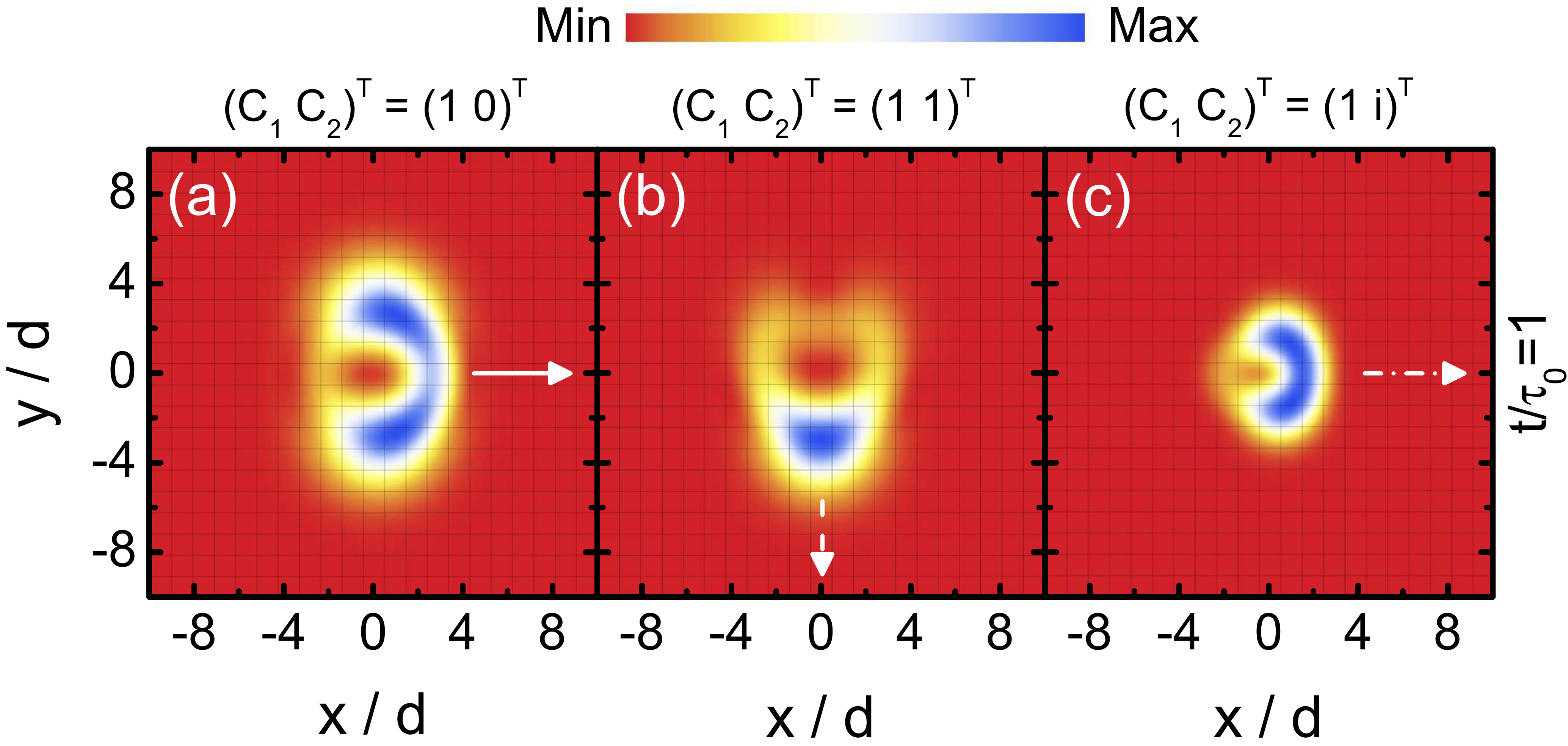}\caption{(Color online) The same as in Fig.~\ref{Fig Mono: Densidade - Monolayer}, but now for BLG and just $t/\tau_{0}=1$.}
\label{Fig.Bi:Densidade - Bilayer}
\end{figure}

For $\left(C_{1}\ C_{2}\right)^{T}=\left(1\ 0\right)^{T}$, the wave packet moves in positive $x$-axis direction and splits in two parts moving along $y$ axis with opposite velocities, Fig.~\ref{Fig.Bi:Densidade - Bilayer}(a). As can be seen from the Fig.~\ref{Fig.Bi:Densidade - Bilayer}(a), the total probability density $\rho\left(x,y,t\right)$ obeys the following symmetry (asymmetry): $\rho\left(x,y,t\right)=\rho\left(x,-y,t\right)$ ($\rho\left(x,y,t\right)\neq\rho\left(-x,y,t\right)$). Therefore, the coordinate $x$ exhibits ZBW. These results are analogous to those in the MLG case (see Sec. \ref{subsec:MONOLAYER-GRAPHENE-CASE}), but with a slightly different deformation shape of the propagated wave function, as illustrated in Fig.~\ref{Fig.Bi:Densidade - Bilayer}.

\begin{figure}[t]
\centering \includegraphics[width=1\columnwidth]{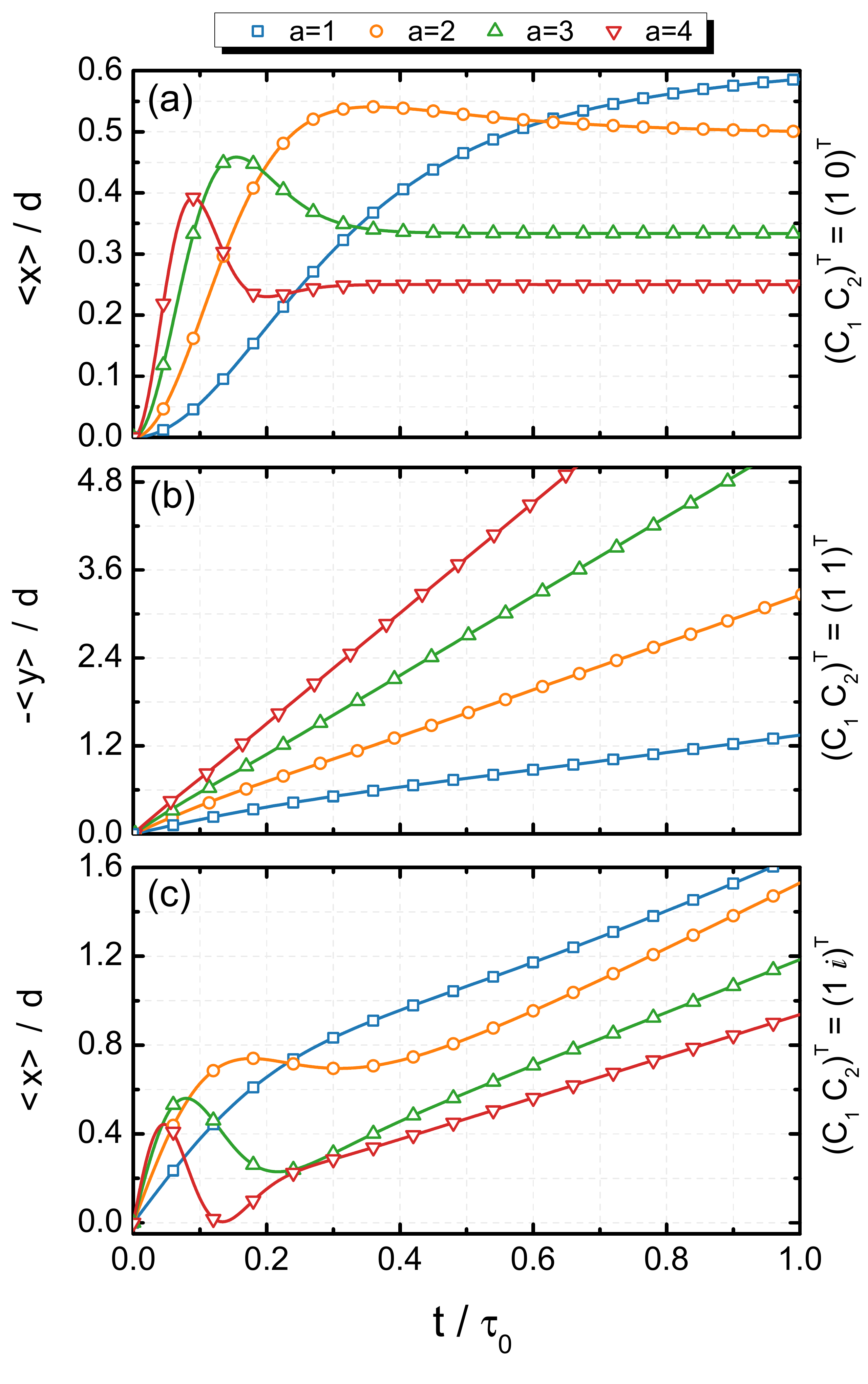}
\caption{(Color online) The same as in Fig.~\ref{Fig Mono: Expectation values}, but now for BLG case
with $\tau_{0}=\gamma d^{2}/\hbar v_{F}^{2}$.}
\label{Fig. Bilayer - Averages}
\end{figure}

Equation (\ref{eq:position operator}) allows us to write the quantities $\left\langle x\right\rangle $ and $\left\langle y\right\rangle $ for BLG as 
\begin{align}
\left\langle x\left(t\right)\right\rangle &\hspace{-0.5mm}=\hspace{-0.5mm}d\hspace{-0.75mm}\left[\hspace{-0.75mm}\frac{1\hspace{-0.75mm}-\hspace{-0.75mm}e^{-a^{2}}}{a}\hspace{-0.75mm}-\hspace{-0.75mm}2e^{-a^{2}}\hspace{-0.75mm}\int_{0}^{\infty}\hspace{-0.5mm}e^{-q^{2}}\text{\text{cos}}\hspace{-0.5mm}\left(2q^{2}t'\right)\hspace{-0.5mm}I_{1}\hspace{-0.5mm}\left(2aq\right)dq\hspace{-0.5mm}\right],\label{eq: Bilayer <x(t)> - (1,0)}
\end{align}
$\left\langle y\left(t\right)\right\rangle =0$, being very similar to the MLG case with the same initial pseudo spin. The analytical (SOT) results for $\left\langle x\left(t\right)\right\rangle $ are illustrated by solid curves (symbols) in Fig.~\ref{Fig. Bilayer - Averages}(a). As shown in Fig.~\ref{Fig. Bilayer - Averages}(a), ZBW has a transient character that is attenuated by an exponential term $e^{-q^2}$ in Eq.~(\ref{eq: Bilayer <x(t)> - (1,0)}) and, after the oscillations disappear, $\left\langle x\left(t\right)\right\rangle /d$ converges to the value of the first term that is time-independent. Unlike the MLG case, Fig.~\ref{Fig Mono: Expectation values}(a), the ZBW frequency for BLG is less affected by increasing $a$.

\subsubsection{$C_{1}=1$ and $C_{2}=1$\label{subsec:BLG - (1,1)}}

The total probability density for $\left(C_{1}\ C_{2}\right)^{T}=\left(1\ 1\right)^{T}$, Fig.~\ref{Fig.Bi:Densidade - Bilayer}(b), obeys the symmetry (asymmetry) relation $\rho\left(x,y,t\right)=\rho\left(-x,y,t\right)$ ($\rho\left(x,y,t\right)\neq\rho\left(x,-y,t\right)$).
Consequently, the $ y $ coordinate is the one that is expected to manifest the ZBW effect. What stands out for this case, is that the wave packet moves along the negative $y-$direction, unlike the MLG case for $\left(1\ 1\right)^{T}$, and does not split into two parts. Its spatial distribution shape and the preferred one-directional propagation ($y$), Fig.~\ref{Fig.Bi:Densidade - Bilayer}(b), seems to be similar to MLG case with pseudospin $\left(1\ i\right)^{T}$, except by the reverse $y$ orientation.

Expectation values of the position $(x,y)$ were obtained in a similar manner as described before and are given by $\left\langle x\left(t\right)\right\rangle =0$ and
\begin{align}
&\left\langle y\left(t\right)\right\rangle\hspace{-0.75mm} =\hspace{-0.75mm}-ae^{-a^{2}}\hspace{-1.5mm}\int_{0}^{\infty}\hspace{-1.5mm}e^{-q^{2}}\hspace{-1.5mm}\left[q\text{sin}\left(2q^{2}t'\right){}_{0}F_{1}\left[3,a^{2}q^{2}\right]\right]\hspace{-0.75mm}dq\nonumber\\
&-4e^{-a^{2}}t'\int_{0}^{\infty}e^{-q^{2}}\left[q^{2}I_{1}\left(2aq\right)+\frac{q}{a}I_{2}\left(2aq\right)\right]dq,\label{eq: Bilayer <y(t)> - (1,1)}
\end{align}
\noindent where $_{0}F_{1}\left[a,z\right]$ in Eq.~(\ref{eq: Bilayer <y(t)> - (1,1)}) is the confluent hypergeometric function. Solid curves (symbols) in Fig.~\ref{Fig. Bilayer - Averages}(b) represent analytical (SOT) results for $\left\langle y\left(t\right)\right\rangle$. As for the MLG case with pseudospin $\left(1\ i\right)^{T}$ (see Fig.~\ref{Fig Mono: Expectation values}(b)), the average position $y$ in the present BLG case exhibits a linear time-dependence with a high group velocity as larger is the $a$ parameter without significant oscillations. It means that ZBW is absent, such that the wave packet in BLG with pseudospin $\left(1\ 1\right)^{T}$ shows to be the appropriated choice in order to investigate transport properties by wave packet dynamics in BLG-based systems within the low-energy approximation described by the two-band model Eq.~(\ref{eq:hamiltonian n-ABC}).

\subsubsection{$C_{1}=1$ and $C_{2}=i$\label{subsec:BLG - (1,i)}}

Assuming $\left(C_{1}\ C_{2}\right)^{T}=\left(1\ i\right)^{T}$, for $t>0$, the wave packet splits into two parts that moves along the $y$-axis in opposite directions, Fig.~\ref{Fig.Bi:Densidade - Bilayer}(c). These two propagating sub-packets with the same probability densities and widths lead to a null average position $\langle y\rangle$ and null expectation value of velocity $\langle v_y\rangle$. As shown in Fig.~\ref{Fig.Bi:Densidade - Bilayer}(c), the probability density $\rho\left(\vec{r},t\right)$ is symmetric (asymmetric) with respect to $y$ ($x$) axis. Due to the lack of mirror symmetry with respect to $x=0$ axis, the wave packet exhibits ZBW along the coordinate $x$, as we had already predicted in Table \ref{Tab: Heisemberg Picture}. It is interesting to note that, if the initial direction of pseudospin coincides with the average momentum $k_{0}^{y}$, for BLG, there is no motion of the wave packet in the $y$-direction, as would be the case for MLG, Sec. \ref{Subsub: Mono C1=00003D1 e C2=00003Di}, but only in the $x$-direction.

By analytically calculating the average value of $x$ and $y$ for this polarization, it leads to 
\begin{multline}
\left\langle x\left(t\right)\right\rangle =de^{-a^{2}}\int_{0}^{\infty}e^{-q^{2}}\left\{ \left[-2\text{sin}\left(2q^{n}t'\right)\right.\right.\\
\left.\left.\cdot\left(-2I_{1}\left(2aq\right)+\frac{2I_{2}\left(2aq\right)}{aq}\right)+\frac{8qtI_{2}\left(2aq\right)}{a}\right]\right\} dq~,\label{eq: Bilayer <x(t)> - (1,i)}
\end{multline}
and $\left\langle y\left(t\right)\right\rangle =0$.
The analytical Green's function based results, obtained by Eq.~(\ref{eq: Bilayer <x(t)> - (1,i)}), are compared to those calculated via SOT within the Dirac model for different parameters $a$, as shown in Fig.~\ref{Fig. Bilayer - Averages}(c). As can be seen in Fig.~\ref{Fig. Bilayer - Averages}(c), there are very similar behaviors with those from MLG case with $\left(C_{1}\ C_{2}\right)^{T}=(1\ 1)^T$, Fig.~\ref{Fig Mono: Expectation values}(c), that is: (i) a transient character of the ZBW, (ii) the $x$ average position is the one that oscillates, (iii) the ZBW amplitude and frequency are directly related to the wave packet width or initial wave vector, such that as higher the parameter $a$, smaller is the oscillation period, vanishing the oscillations faster in time and converging the group velocity $v_x$ to a constant non-zero value.

\subsection{ZBW in TLG\label{subsec:trilayer-GRAPHENE-CASE}}

As the last example of our investigations on ZBW in NLG, we studied the dynamics of wave packet in ABC-stacked TLG, as illustrated in Fig. \ref{Fig: Geometries-1}. Expectation values of $x$ and $y$ coordinates as a function of time are obtained with the same analytical and numerical methods used so far, therefore, details of these calculations for TLG will be omitted.

\begin{figure}[t]
\centering \includegraphics[width=1\columnwidth]{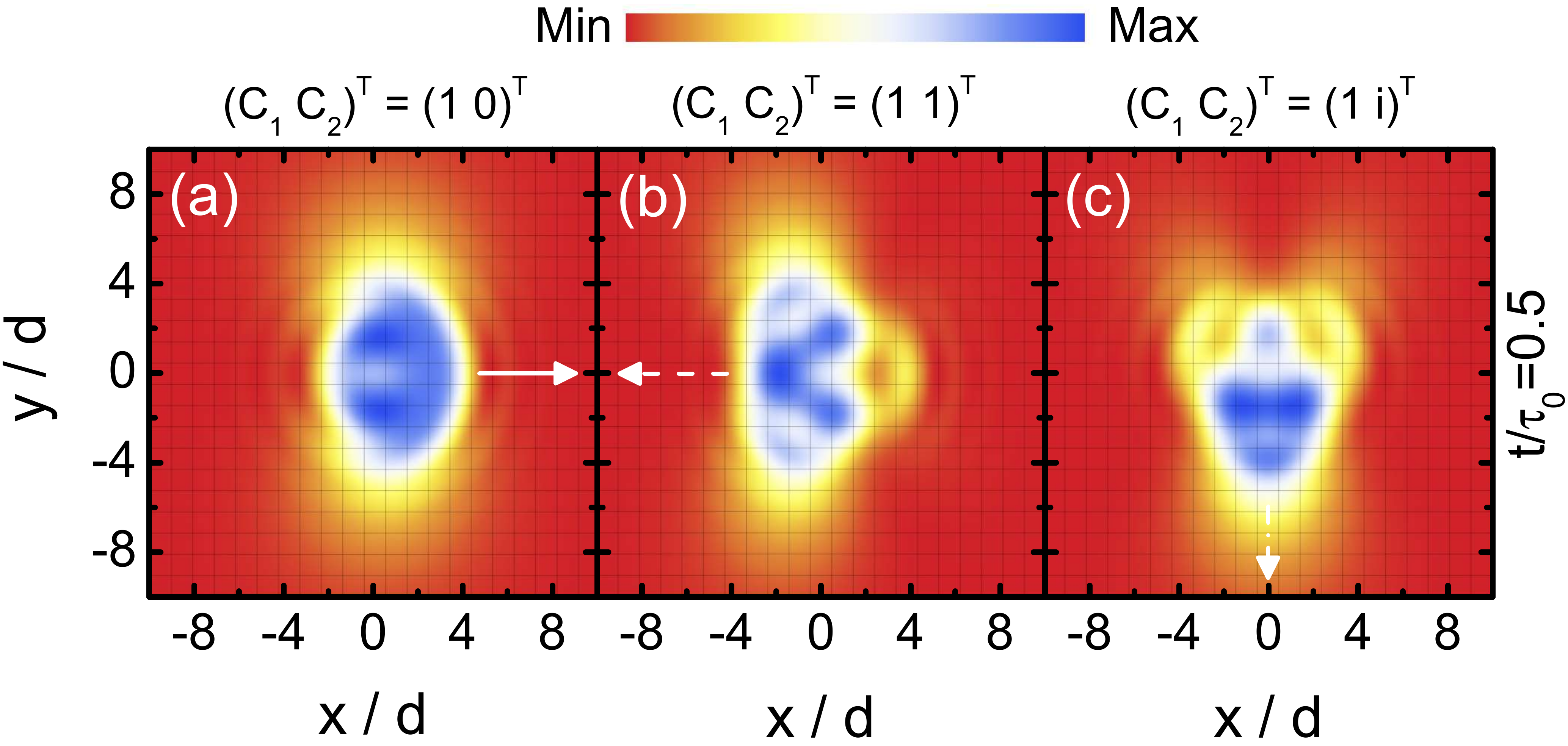}\caption{(Color online) The same as in Fig.~\ref{Fig Mono: Densidade - Monolayer}, but now for TLG at $t/\tau_{0}=0.5$.}
\label{Fig. Densidade - Trilayer}
\end{figure}

\begin{figure}[!t]
\centering \includegraphics[width=1\columnwidth]{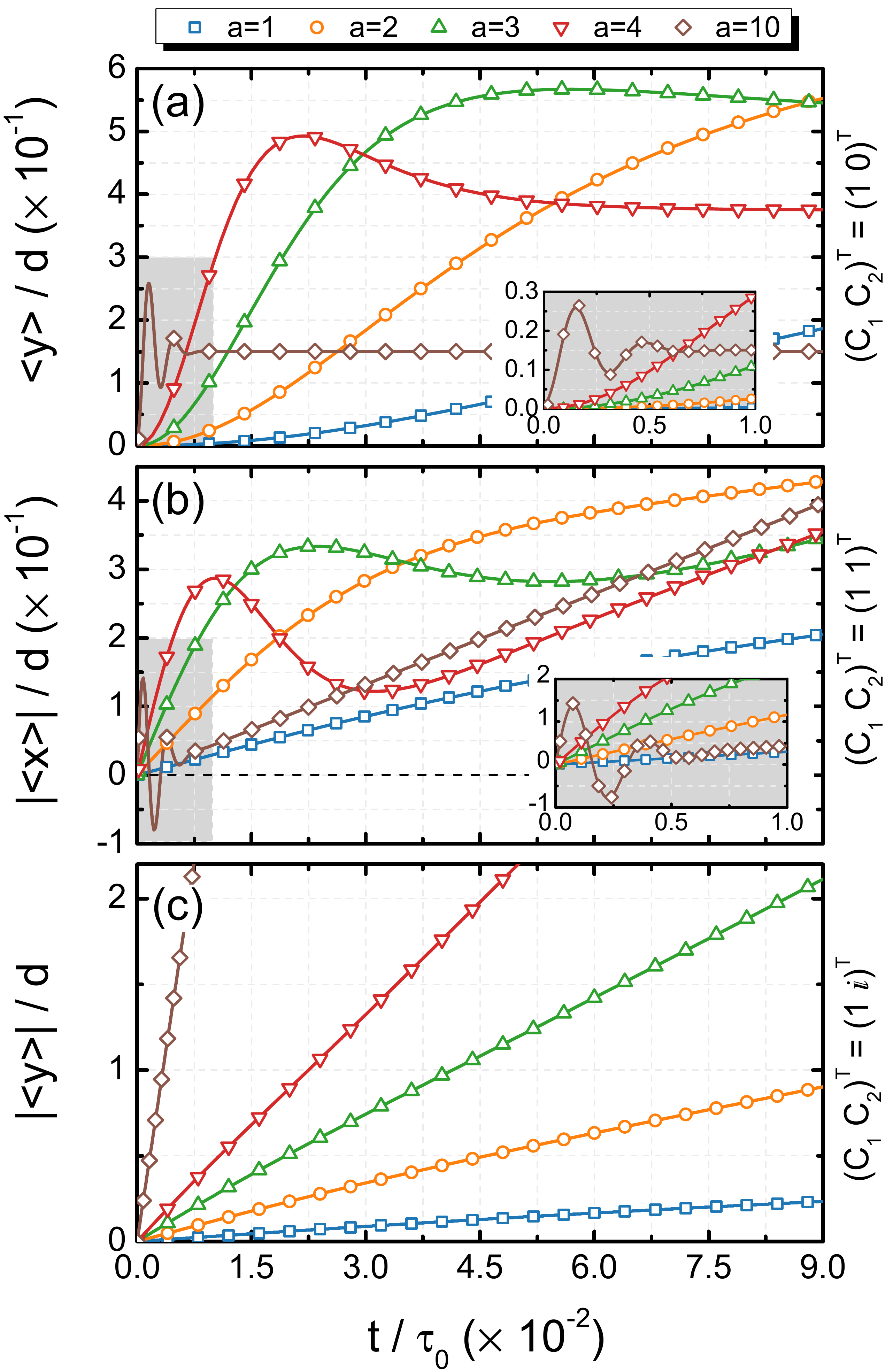}
\caption{(Color online) The same as in Fig.~\ref{Fig Mono: Expectation values} (including $a=10$), but now for TLG with $\tau_{0}=\gamma^{2}d^{3}/\hbar^{2}v_{F}^{2}$. The insets in panels (a) and (b) show magnification of the gray shaded areas for better visualization at small $t/\tau_0$ values.}

\label{Trilayer - All cases - Average}
\end{figure}

Assuming $\left(C_{1}\ C_{2}\right)^{T}=\left(1\ 0\right)^{T}$, one obtains
\begin{align}
&\left\langle x\left(t\right)\right\rangle \hspace{-1mm}=\hspace{-1mm}3d\hspace{-1mm}\left(\hspace{-1mm}\frac{1\hspace{-1mm}-\hspace{-1mm}e^{-a^{2}}}{2a}\hspace{-1mm}\right)\hspace{-1mm}-\hspace{-1mm}3de^{-a^{2}}\hspace{-2mm}\int_{0}^{\infty}\hspace{-3mm}e^{-q^{2}}\text{\text{cos}}\left(2q^{3}t'\right)\hspace{-1mm}I_{1}\left(2aq\right)dq,\label{eq: <x(t)> - (1,0) - trilayer}
\end{align} and $\left\langle y\left(t\right)\right\rangle=0$. The probability density and a comparison between the analytical results, Eq.~(\ref{eq: <x(t)> - (1,0) - trilayer}), and those from SOT within the Dirac model, are represented in Fig.~\ref{Fig. Densidade - Trilayer}(a) and Fig.~\ref{Trilayer - All cases - Average}(a) for different parameters $a$ as a function of time. As $a$ increases, Fig.~\ref{Trilayer - All cases - Average}(a), the ZBW becomes more evident, although still exhibiting a transient character, as in the previous MLG and BLG cases.
On the other hand, for the pseudospin configuration $\left(C_{1}\ C_{2}\right)^{T}=\left(1\ 1\right)^{T}$
the results for expectation value of the position of the wave packet
are given by
\begin{align}
&\left\langle x\left(t\right)\right\rangle =-\frac{3de^{-a^{2}}}{2a^{2}}\int_{0}^{\infty}\frac{e^{-q^{2}}}{q^{2}}\left\{6aq^{4}I_{3}(2aq)t\right.\nonumber\\
&\left.+\left[\left(2a^{2}q^{2}+6\right)I_{2}(2aq)-3aqI_{1}(2aq)\right]\sin\left(2q^{3}t\right)\right\}dq
\label{eq: <x(t)> - (1,1) - trilayer}
\end{align}
\noindent and $\left\langle y\left(t\right)\right\rangle =0$. Figure \ref{Trilayer - All cases - Average}(b) shows $\left\langle x\left(t\right)\right\rangle$, Eq.~(\ref{eq: <x(t)> - (1,1) - trilayer}), and the SOT results calculated within the Dirac model. As we can be seen in Fig.~\ref{Trilayer - All cases - Average}(b) and its inset with an enlargement for small time steps, after the transient oscillatory behaviour, $\left|\left\langle x\right\rangle \right|$ increases linearly with time converging to a non-null constant group velocity $v_x$ in a similar way as observed for MLG case with pseudospin $(1\ 1)^T$ (see Fig.~\ref{Fig Mono: Expectation values}(b)) and for BLG case with pseudospin $(1\ i)^T$ (see Fig.~\ref{Fig. Bilayer - Averages}(c)). The probability density is illustrated in Fig.~\ref{Fig. Densidade - Trilayer}(b) and shows that the direction of propagation of the wave packet is in accordance with Eq.~(\ref{eq: <x(t)> - (1,1) - trilayer}).
\begin{figure}[t]
\centering \includegraphics[width=1\columnwidth]{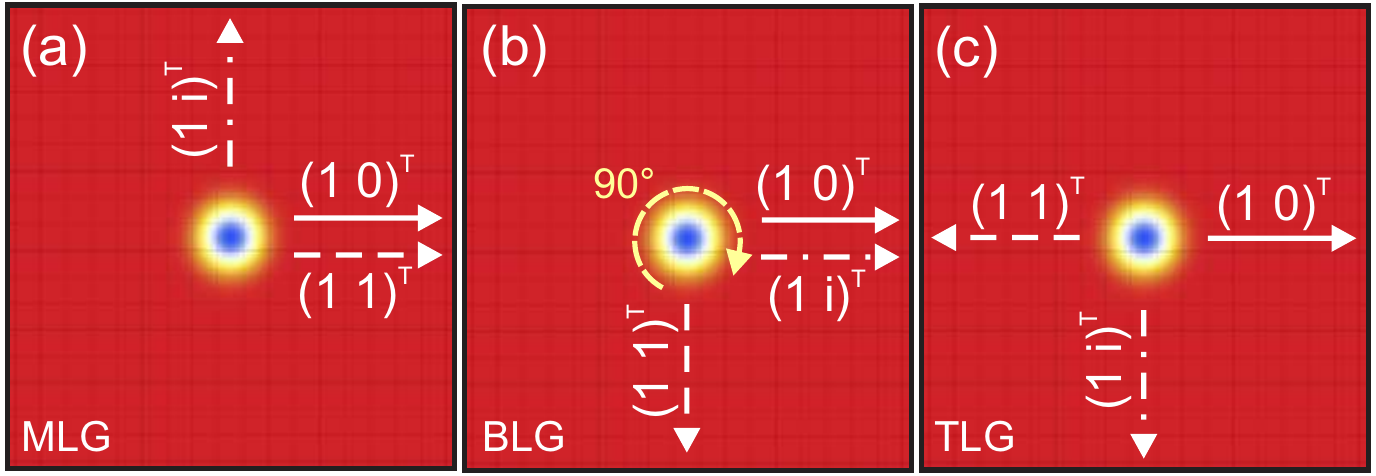}\caption{(Color online) Representation of the different directions of propagation of the Gaussian wave packet according to the choice of initial pseudospinor for (a) MLG, b) bilayer and c) trilayer graphene, obtained from Eq. (\ref{eq:two components}). The solid, dashed and dash-dotted white curves represent the initial pseudospinor defined as $\left(C_{1}\ C_{2}\right)^{T}=\left(1\ 0\right)^{T}$, $\left(C_{1}\ C_{2}\right)=\left(1\ 1\right)^{T}$ and $\left(C_{1}\ C_{2}\right)=\left(1\ i\right)^{T}$, respectively. The long-dashed circle in (b) indicates that when one includes one more layer the direction of propagation of the wave packet motion rotates by $90\lyxmathsym{\textdegree}$ for the pseudospinor $\left(1\ 1\right)^{T}$ and $\left(1\ i\right)^{T}$.}
\label{Monolayer-Bilayer-Trilayer-Direction}
\end{figure}

Finally, for the pseudospinor $(1\ i)^T$ the expectation values of the position operator are $\left\langle x\right\rangle =0$ and
\begin{align}
&\left\langle y\right\rangle =\frac{-3e^{-a^{2}}}{2a^{2}}\int_{0}^{\infty}\frac{e^{-q^{2}}}{q}\left(4q^{2}t\left(a^{2}q^{2}+3\right)I_{2}(2aq)\right.\nonumber\\
&\hspace{1cm}\left.-6aq^{3}tI_{1}(2aq)+3aI_{3}(2aq)\sin\left(2q^{3}t\right)\right)dq.
\label{eq:<y(t)> - (1,i) - trilayer}
\end{align}

Figure \ref{Trilayer - All cases - Average}(c) provides a comparison between the analytical results, obteined numerically from Eq.~(\ref{eq:<y(t)> - (1,i) - trilayer}), with those obtained by the SOT within the Dirac model. This results shows to be analogous to the MLG case for $\left(C_{1}\ C_{2}\right)^{T}=\left(1\ i\right)^{T}$ and BLG case for $\left(C_{1}\ C_{2}\right)^{T}=\left(1\ 1\right)^{T}$, where (i) ZBW is absent; and (ii) as $a$ increases, $\left\langle y\right\rangle /d$ also increases linearly with time without visible oscillations and with a non-null constant group velocity along $y-$direction.

\subsection{Influence of the number of graphene layers on wave packet dynamics \label{subsec:DYNAMICS}} 

As observed in Secs. \ref{subsec:MONOLAYER-GRAPHENE-CASE}, \ref{subsec:BILAYER-GRAPHENE-CASE} and \ref{subsec:trilayer-GRAPHENE-CASE}, for different pseudospin polarization $\left(C_{1}\ C_{2}\right)^{T}=\left(1\ 1\right)^{T}$ and $\left(C_{1}\ C_{2}\right)^{T}=\left(1\ i\right)^{T}$, the wave packet exhibits different propagation directions for MLG, BLG and TLG (for more details, see Appendix II). Figure \ref{Monolayer-Bilayer-Trilayer-Direction} illustrates these three situations. In fact, such change in propagation direction is expected as $n$ increases, since the low-energy Hamiltonian for $ABC$-NLG has Pauli matrices $\sigma_{x}$ and $\sigma_{y}$ multiplying both $k_{x}$ and $k_{y}$ for $n\geq2$, unlike the MLG case. For example, for BLG, $H_{BLG}=\hbar^2 v_F^2\gamma^{-1}\left[\left(k_{x}^{2}-k_{y}^{2}\right)\sigma_{x}+2k_{x}k_{y}\sigma_{y}\right]$. Consequently, the velocity components in $x-$ and $y-$directions, calculated according to the steps in Sec. \ref{subsec:Heisenberg-Picture}, are expected to be proportional to $2\hbar v_{F}^{2}\gamma^{-1}k_{y}\langle\sigma_{y}\rangle$ and $-\hbar v_{F}^{2}\gamma^{-1}k_{y}\langle\sigma_{x}\rangle$, respectively, where we already took into account that the wave packet momentum in Eq. (\ref{eq:initial wave}) has only a component in the $y$-direction, i.e. $k_x \equiv 0$. As for TLG, the same procedure leads to velocity components in $x-$ and $y-$directions proportional to $\ensuremath{-3\hbar^{2}v_{F}^{3}\gamma^{-2}k_{y}^{2}\langle\sigma_{x}\rangle}$ and $\ensuremath{-\hbar^{2}v_{F}^{3}\gamma^{-2}k_{y}^{2}\langle\sigma_{y}\rangle}$, respectively. Thus, for a given initial pseudospin orientation, these expressions help to qualitatively predict the observed changes in propagation direction and the increasing propagation velocity as the number of layers increases, whereas the detailed behavior of the wave packet dynamics and its ZBW requires the more sophisticated approaches described in the previous Sections. Moreover, by comparing the transient duration time ($t^d$) in Figs.~\ref{Fig Mono: Expectation values}, \ref{Fig. Bilayer - Averages} and \ref{Trilayer - All cases - Average} and the wave packet evolution in Figs.~\ref{Fig Mono:  Densidade - Monolayer}, \ref{Fig.Bi:Densidade - Bilayer} and \ref{Fig. Densidade - Trilayer} for MLG, BLG and TLG, respectively, one can realize that as the number of layers increases, the propagating wave function spreads faster for a certain fixed time range, that in turn leads to smaller time scales for the transient behavior, i.e. $t^d_{N=3}<t^d_{N=2}< t^d_{N=1}$.

\subsection{Dirac valley selection for wave packet dynamics}

The choice of the propagation direction in real space also depends on which Dirac valley the initial wave packet is taken, since the $k_x^{D}$ and $k_y^{D}$ directions in the Dirac model are rotated with respect to the $k_x^{TB}$ and $k_y^{TB}$ tight-binding directions via the standard 2D rotation matrix:
\begin{equation}
\left(\hspace{-0.15cm}\begin{array}{c}
k_x^{D}\\
k_y^{D}
\end{array}\hspace{-0.15cm}\right)=\left(\hspace{-0.15cm}\begin{array}{cc}
\cos\theta & -\sin\theta\\
\sin\theta & \cos\theta
\end{array}\hspace{-0.15cm}\right)\left(\hspace{-0.15cm}\begin{array}{c}
k_x^{TB}\\
k_y^{TB}
\end{array}\hspace{-0.15cm}\right),\label{eq.rotation}
\end{equation} with $\theta = \pi/2$, $7\pi/6$, and $11\pi/6$ [$\theta = \pi/6$, $5\pi/6$, and $3\pi/2$] for $K$ ($K'$) Dirac valleys of the first Brillouin zone. In addition, since in our analysis the time-reversal symmetry is preserved, then $H(\vec{k})=H(-\vec{k})^*$ and the low-energy bands are doubly degenerate. As a consequence, all results obtained along this work for $K$ Dirac valley can be easily mapped into the $K'$ valley by just rotating the reciprocal space vectors according to Eq.~(\ref{eq.rotation}).

\section{Conclusions\label{sec:Conclusions}}

A comprehensive study of the quantum dynamics of charged particles represented by a 2D Gaussian wave packet in multilayer graphene has been presented. Using the Green's function method, we obtained generalized analytical expressions for the time dependence of the wave functions in $ABC$-stacked NLG that allowed us to calculate the average values of position operators for an arbitrary number of graphene layers $n$.

A semi-analytical method, which allows one to calculate wave packet scattering by arbitrary potential profiles is proposed. The method is based on the well-known SOT, adapted here for the $2\times 2$ Dirac approximation for the multi-layer graphene Hamiltonian. Analytical results for the expectation values of the position of the center of the wave packet show perfect agreement with those from the SOT within the Dirac approximation, for all cases of initial pseudospin orientation investigated here. This consolidates the methods proposed here, which are suitable for large graphene samples with any number of ABC-stacked layers (in contrast to tight-binding models, where the computational cost rapidly increases with the number of atoms), as very useful tools for continuum model investigations of transport properties in multilayer graphene. 

As examples, the proposed methods here are applied to the study of the dynamics of wave packets in $ABC$-stacked MLG, BLG and TLG, with different pseudospin polarization. Our results demonstrate how ZBW depends on the number of graphene layers. Wave packets with the same pseudospin orientation in MLG, BLG and TLG are shown to propagate in different directions and with different velocities. ZBW is shown to be minimized as the pseudospin orientation is taken the same as the wave packet momentum. For the parameters considered in this paper, when both the pseudospin and momentum are oriented along the $y$-direction (i.e. assuming $\langle \sigma_y \rangle \neq 0$, $(C_1\ C_2)^T = (1\ i)^T$, $p_{0y}\neq 0$ and $k_x \equiv 0$), the wave packet position is approximately a linear function of time, propagating along the $+y$-, $+x$- and $-y$-directions for MLG, BLG, and TLG, respectively. The ZBW phenomena in multilayer graphene displays a transient behavior, i.e. the oscillations of the physical observables decay with time and a natural damping is observed. Our results show that the transient behavior time $t^d$ is of the order of dozens of femtoseconds and the larger the number of layers the shorter the transient time, i.e. $t^d_{N}<t^d_{N-1}$. At the experimental point-of-view, the amplitude of the oscillations should depend very strongly on the duration of the applied pulse, whereas the duration time of the total damping is due to the light emission time scale. The latter condition is owing to the fact that the electron oscillations give rise to a time-dependent dipole moment which will be a source of electric field and it will emit or absorb radiation in the far infrared range \cite{Rusin2009, rusin2014zitterbewegung}.

Both theoretical methods proposed here will be useful for future simulations of wave packet propagation and scattering in multilayer graphene, and that the discussions about the results found in this work will contribute to a better understanding of ZBW in these systems.

\section*{ACKNOWLEDGMENTS}
Discussions with D. J. P. de Sousa and J. M. Pereira Jr. are gratefully acknowledged. This work was financially supported by the Brazilian Council for Research (CNPq), under the PQ and PRONEX/FUNCAP programs, and by CAPES. One of us (B. V. D.) is supported by the FWO-Vl. D.R.C is supported by CNPq grant numbers 310019/2018-4 and 437067/2018-1.

\section*{Appendix I: Wave functions in terms of the Bessel function}\label{Appendix I}

Using cylindrical coordinates in order to rewrite Eqs. (\ref{eq:PHI1 PHI4}) and (\ref{eq:PHI2 PHI3}) in
terms of Bessel Functions, Eqs. (\ref{eq:PHI1 PHI4-1}) and (\ref{eq:PHI2 PHI3-1}), the following variable substitutions
need to be made: $a=k_{0}d$, $q=pd/\hbar$ and $\vec{p}=\left(p\text{cos}\phi,p\text{sen}\phi\right)$.
On the other hand, for convenience, but without loss of generality, we introduce the following dimensionless variables: 
\begin{equation}
t\hspace{0.2cm}\rightarrow\hspace{0.2cm}t'=\frac{\hbar^{n-1}t}{\gamma d^{n}},\label{eq:108}
\end{equation}
where $n$ is the number of layers, $x\rightarrow x'=x/d$, $y\rightarrow y'=y/d$
and $r\rightarrow r'=r/d$.
Easily we get $p^{n}t/\gamma\hbar\rightarrow q^{n}t$, $-\left(k_{0}^{y}d\right)^{2}/2\rightarrow-a^{2}/2$
and, consequently,
\begin{equation}
i\frac{\vec{p}\cdot\vec{r}}{\hbar}\hspace{-0.05cm}-\hspace{-0.05cm}\frac{p^{2}d^{2}}{2\hbar^{2}}\hspace{-0.05cm}-\hspace{-0.05cm}\frac{p_{y'}k_{0}d^{2}}{\hbar}\hspace{-0.05cm}=\hspace{-0.05cm}iq\left(x'\text{cos}\phi\hspace{-0.05cm}+\hspace{-0.05cm}y'\text{sen}\phi\right)\hspace{-0.05cm}-\hspace{-0.05cm}\frac{q^{2}}{2}\hspace{-0.05cm}+\hspace{-0.05cm}qa\text{sen}\phi,
\end{equation}
which are the argument of sine (cosine) and the two exponential in Eqs.~(\ref{eq:PHI1 PHI4}) and (\ref{eq:PHI2 PHI3}).
Now, from the fact that $d\vec{p}=\left(\hbar^{2}/d^{2}\right)qdqd\phi$, the integral in $d\vec{p}$ in Eqs.~(\ref{eq:PHI1 PHI4}) and (\ref{eq:PHI2 PHI3}) can be splitted into two others, as follows:
\begin{equation}
\int_{-\infty}^{+\infty}d\vec{p}\hspace{0.2cm}\rightarrow\hspace{0.2cm}\int_{0}^{\infty}qdq\int_{-\pi}^{+\pi}d\phi.\label{eq:128-1}
\end{equation}
Therefore, after replacing the transformation aforementioned and solving the integrals in $\phi$, we obtain the two components $\Phi_{1}\left(\vec{r},t\right)$ and $\Phi_{2,3}\left(\vec{r},t\right)$, Eqs. (\ref{eq:PHI1 PHI4-1}) and (\ref{eq:PHI2 PHI3-1}), respectively, of the wave function in terms of the integral in $dq$ and the Bessel functions.

\section*{Appendix II: Direction of the wave packet as a function of layers number} \label{Appendix II}

Analytically, a general expression for $\left\langle x\left(t\right)\right\rangle$ and $\left\langle y\left(t\right)\right\rangle$, in cylindrical coordinate, as a function of N-ABC layers, can be obtained from Eq.~(\ref{eq:position operator}). Since for $\left(C_{1}\ C_{2}\right)^{T}=\left(1\ 0\right)^{T}$ the wave packet always moves in the positive direction of the $x$-axis, as shown in Fig.~\ref{Monolayer-Bilayer-Trilayer-Direction}, we analysed here only the other two initial pseudospinor configuration, i.e $(1\ 1)^{T}$ and $(1\ i)^{T}$. Thus, for these cases, $\left\langle x\left(t\right)\right\rangle $ is defined, respectively, as
\begin{subequations}
\begin{equation}
\left\langle x\left(t\right)\right\rangle \hspace{-0.05cm}=\hspace{-0.05cm}\alpha\hspace{-0.05cm}\left(2q^{n}t\cos\hspace{-0.05cm}\left(\phi\right)\cos\hspace{-0.05cm}\left(n\phi\right)\hspace{-0.05cm}+\hspace{-0.05cm}\sin\hspace{-0.05cm}\left(2q^{n}t\right)\sin\hspace{-0.05cm}\left(n\phi\right)\sin\hspace{-0.05cm}\left(\phi\right)\right)
\label{Eq. <x> as function of N (1 1)}
\end{equation}
\begin{equation}
\left\langle x\left(t\right)\right\rangle \hspace{-0.05cm}=\hspace{-0.05cm}\ensuremath{\alpha}\hspace{-0.05cm}\left(2q^{n}t\cos\hspace{-0.05cm}\left(\phi\right)\sin\hspace{-0.05cm}\left(n\phi\right)\hspace{-0.05cm}-\hspace{-0.05cm}\sin\hspace{-0.05cm}\left(2q^{n}t\right)\cos\hspace{-0.05cm}\left(n\phi\right)\hspace{-0.05cm}\sin\left(\phi\right)\right),
\label{Eq. <x> as function of N (1 i)}
\end{equation}
\end{subequations}
where $\alpha=\left(nde^{-a^{2}}/2\pi\right)\int e^{-q^{2}}dq\int e^{2aq\sin(\phi)}d\phi$. Solving the integral in $\phi$ of Eq.~(\ref{Eq. <x> as function of N (1 1)}) (Eq.~(\ref{Eq. <x> as function of N (1 i)})), we concluded that for n even (odd), $\left\langle x\left(t\right)\right\rangle $ is null. On the other hand, the opposite occurs for $\left\langle y\left(t\right)\right\rangle $ (this can be verified in a similar way). This alternation of the nullity of $\left\langle x\left(t\right)\right\rangle $ and $\left\langle y\left(t\right)\right\rangle$, for up to 3 layers, for different initial pseudospinor, are illustrated in Fig.~\ref{Monolayer-Bilayer-Trilayer-Direction}.

\bibliographystyle{apsrev4-2}
\bibliography{myreferences}

\end{document}